\documentclass{article}
\usepackage[utf8]{inputenc}
\usepackage{authblk}
\usepackage{setspace}
\usepackage[margin=1.25in]{geometry}
\usepackage{graphicx}
\graphicspath{ {./figures/} }
\usepackage{subcaption}

\usepackage{amsmath}
\usepackage{textcomp}
\usepackage{newunicodechar}
\usepackage{amsmath,amssymb}
\usepackage{upgreek}
\newunicodechar{−}{\textminus}

\usepackage[style=nejm, 
citestyle=numeric-comp,
sorting=none]{biblatex}
\title{Influence of Magnetospheric Plasma Environment on Surface Charging of the Lunar South Pole}

\author[1]{Yaqi Xiao}
\author[1]{Xiaoman Wang}
\author[1*]{Ronghui Quan}

\affil[1]{College of Astronautics, Nanjing University of Aeronautics and Astronautics, Nanjing, China.}
\affil[*]{Address correspondence to: quanrh@nuaa.edu.cn}

\date{}

\begin{document}

\maketitle

\begin{abstract}
The lunar south pole is a key candidate region for future lunar exploration and base construction, but its charging characteristics under real topographic conditions and dynamic plasma environments remain insufficiently understood. A high-fidelity terrain model of the lunar south pole spanning 86°S-90°S was constructed from optimized LRO/LOLA elevation data. Surface charging evolution over half a lunar orbital cycle was then simulated with a finite element-BP neural network scheme, using lunar-phase-dependent plasma inputs encompassing plasma parameters of solar wind and diverse Earth magnetospheric zones. The results show that south polar topography strongly regulates surface charging. Higher potentials appear on windward terrains, whereas lower potentials occur in shielded leeward regions, leading to enhanced local electric fields at the tops of uplands and crater floor-wall boundaries. Significant potential differences between the crests and the middle of the downstream walls of various craters indicate that these regions are highly terrain-sensitive. When the Moon passes through Earth's magnetosphere, surface potential and electric field are roughly symmetric around 0° lunar phase. From the solar wind to the plasma sheet, surface potential generally decreases while electric field magnitude rises. Only in the narrow magnetotail lobe adjacent to the plasma sheet does the potential temporarily increase and the electric field weaken. In the plasma sheet, the surface potential can decrease to approximately -1000 V, and the domain's peak electric field reaches about 5 V/m. These findings provide references for landing site selection, rover path planning, and electrostatic protection of lunar surface equipment.
\end{abstract}


\section{Introduction}

\begin{figure}[h]
    \centering
    \includegraphics[width=0.5\textwidth]{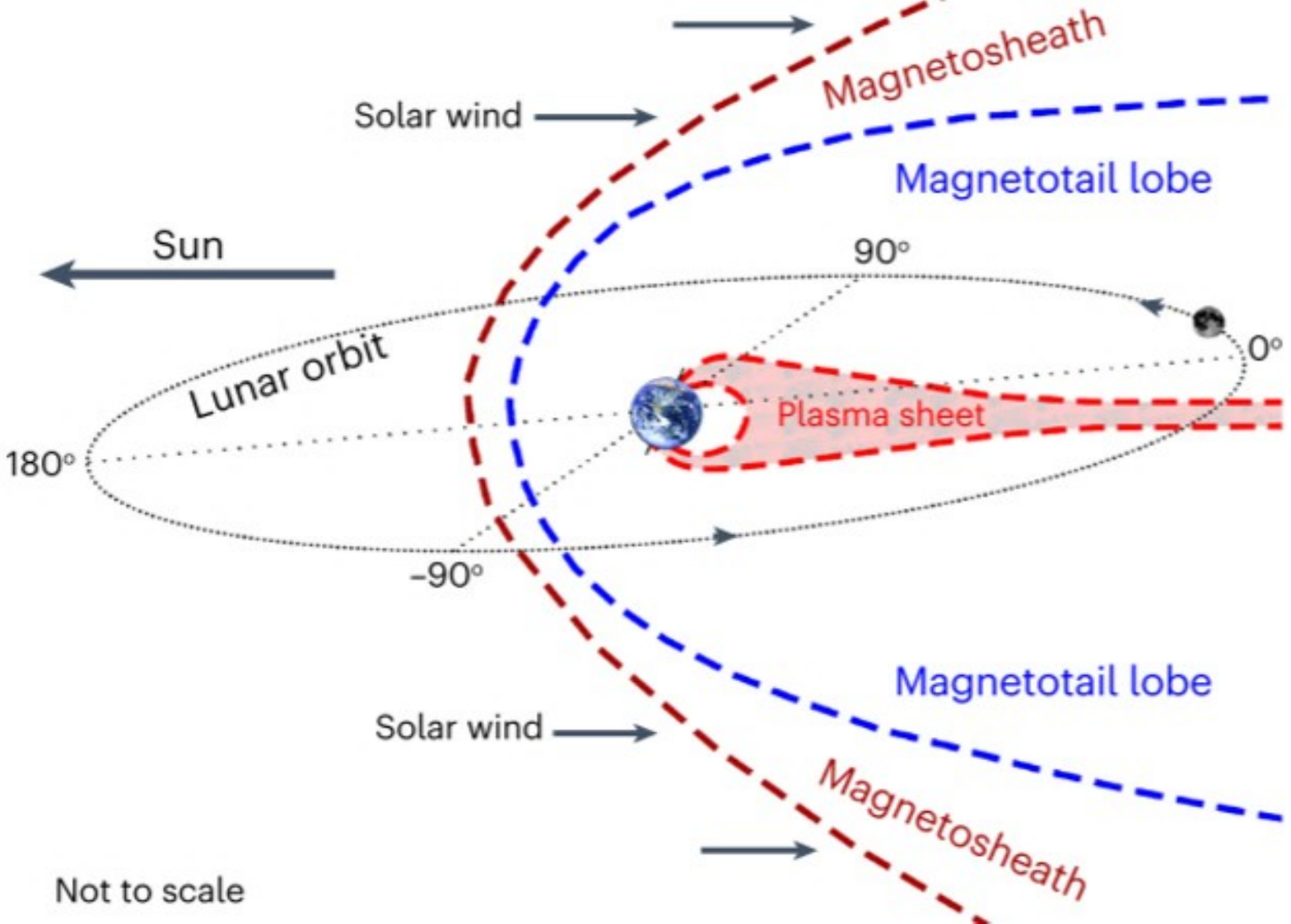}
    \caption{Side-view schematic diagram showing the configuration of the magnetotail.\cite{LiS1}.}
    \label{fig:1}
\end{figure}

The Moon has neither a dense atmosphere nor an intrinsic global magnetic field. As a result, its surface is directly exposed to solar radiation, the solar wind, and magnetospheric plasma, which leads to surface charging processes\cite{Halekas1,Poppe1,Harada1,Nishino1}. Under the combined effects of photoelectron emission, electron and ion deposition, and secondary electron emission, complex surface potential distribution and near-surface electric field can be formed on the lunar surface\cite{Farrell2,Halekas2,Stubbs1}. During its orbital motion, the Moon spends most of the time in the solar wind, but it also periodically passes through different regions of the Earth’s magnetosphere, as illustrated in Figure \ref{fig:1}. 

The plasma characteristics in these regions differ significantly. For example, the magnetosheath generally has a relatively high plasma density, whereas the magnetotail lobe and plasma sheet has lower plasma densities and higher particle energies\cite{Halekas3,Halekas4,Harada2,Poppe2}. These variations in the plasma environment can lead to distinct lunar surface charging behaviors. Lunar surface charging not only affects the charging and transport of lunar dust, but may also threaten the operational safety of landers, probes, lunar surface facilities, and future lunar bases.

In recent years, many countries have been actively preparing for long-term lunar exploration missions and planning to establish lunar bases. The lunar south polar region has become a major candidate area for lunar base site selection, as its craters may contain permanently illuminated peaks on the rims and potential water-ice deposits on the floors\cite{Wang1,Farrell3}. Meanwhile, the dense distribution of craters creates highly complex terrain, and the motion of plasma in this region is strongly affected by solar wind orographic effects\cite{Farrell1}. Specifically, plasma flow is blocked by elevated landforms such as upstream crater walls and mountains, generating complex plasma wake downstream and further inducing a complicated electrostatic environment on the lunar surface. Therefore, investigating the surface charging characteristics of the lunar south polar region under different space plasma environments is of great significance to lunar exploration missions and long-term human presence on the Moon.

Numerical calculation and simulation of the Moon's global surface charging are well-developed, and many scholars have further studied lunar surface charging under complex terrains. Farrell et al.\cite{Farrell1} used an analytical model to investigate the electrical environment of permanently shadowed craters near the lunar south pole. Terrain including mountains and crater walls obstructs the solar wind and forms plasma void on the leeward side. Electrons, with smaller mass and higher speed, move into shadowed areas faster than ions, resulting in leeward potentials below -50 V. Zimmerman et al.\cite{Zimmerman1} adopted ideal stepped crater models and found that the wake structure is mainly controlled by four dimensionless parameters: wake inflow angle, effective obstruction size, normalized electron impact energy and secondary electron yield. Solar storms change local parameters and thus affect particle flux, bipolar fields and wake potentials. Lund et al.\cite{Lund1} established geometric models of small craters based on lunar crater morphological parameters. Craters produce local plasma wake, with the leeward surface potential reaching −35.2 V. The electric field in regolith depends on charge deposition, regolith permittivity and layer thickness. The charging state of lunar modules varies with their orientation relative to craters and solar wind. Using the SPIS-DUST code, Qing Xia et al.\cite{Xia1} simulated a 5 m-high crater wall. Terrain blocking creates quasi-neutral conditions in the shadowed plasma void, where dust density is one order of magnitude lower than elsewhere, while intense dust transport appears near the terminator. Hong Gan et al.\cite{Gan1} employed a 5 m-diameter hemisphere to represent a crater and simulated the charging state of the Chang'e-7 rover at various positions near the crater under typical slow solar wind conditions. Stronger rim shielding and wake effects on the leeward side produce lower potentials, exposing the rover to greater risks of severe negative charging during operations. Chengxuan Zhao et al.\cite{Zhao1} divided the surface of Shackleton Crater into three zones: sunlit area, leeward electron cloud region and downstream region. Analytical calculation shows that the leeward potential drops to −175 V. They further adopted SPIS-DUST to simulate a crater model scaled down by 1000 times for verification. The results indicate that the surface potential inside the crater is below −30 V, and the number density of lunar dust near crater walls reaches $10^5\ \mathrm{m}^{-3}$. 

However, most of these studies adopted oversimplified, small-scale crater geometric models with fixed plasma environment inputs. Few investigations have accounted for the actual lunar topography and realistic plasma conditions. In this study, we simulate the lunar surface charging process over the south polar region using the finite element method. The terrain model is built upon NASA topographic survey data, and a trained BP neural network is applied to compute surface potentials for higher computational efficiency. First, the influence of varying solar elevation angles on charging results is analyzed under typical solar wind conditions. Then, with the solar elevation angle set to 1°, dynamic plasma parameters are adopted to simulate variations in surface charging as the Moon passes through the solar wind, magnetosheath, magnetotail lobe, and plasma sheet over half of its orbital period. This work can provide theoretical references for future lunar exploration and lunar base construction.

Chapter 2 elaborates on model establishment and parameter configuration. Chapter 3 presents and analyzes the simulation results, and Chapter 4 concludes the entire study.

\section{Methods}
\subsection{Lunar South Polar Terrain Model}

To investigate the surface charging characteristics of the lunar south pole under realistic topographic conditions, this study adopts the digital elevation model (DEM) optimized by Barker et al. as the topographic dataset\cite{Barker1}. This model covers the region from 80°S to 90°S with a pixel resolution of 80 meters per pixel (MPP). The original data were derived from observations by the Lunar Orbiter Laser Altimeter (LOLA) onboard the Lunar Reconnaissance Orbiter (LRO), which guarantees high topographic accuracy and a reliable geodetic reference frame. 

\begin{figure}[h]
    \centering
    \begin{subfigure}{0.4\textwidth}
        \includegraphics[height=1.8in]{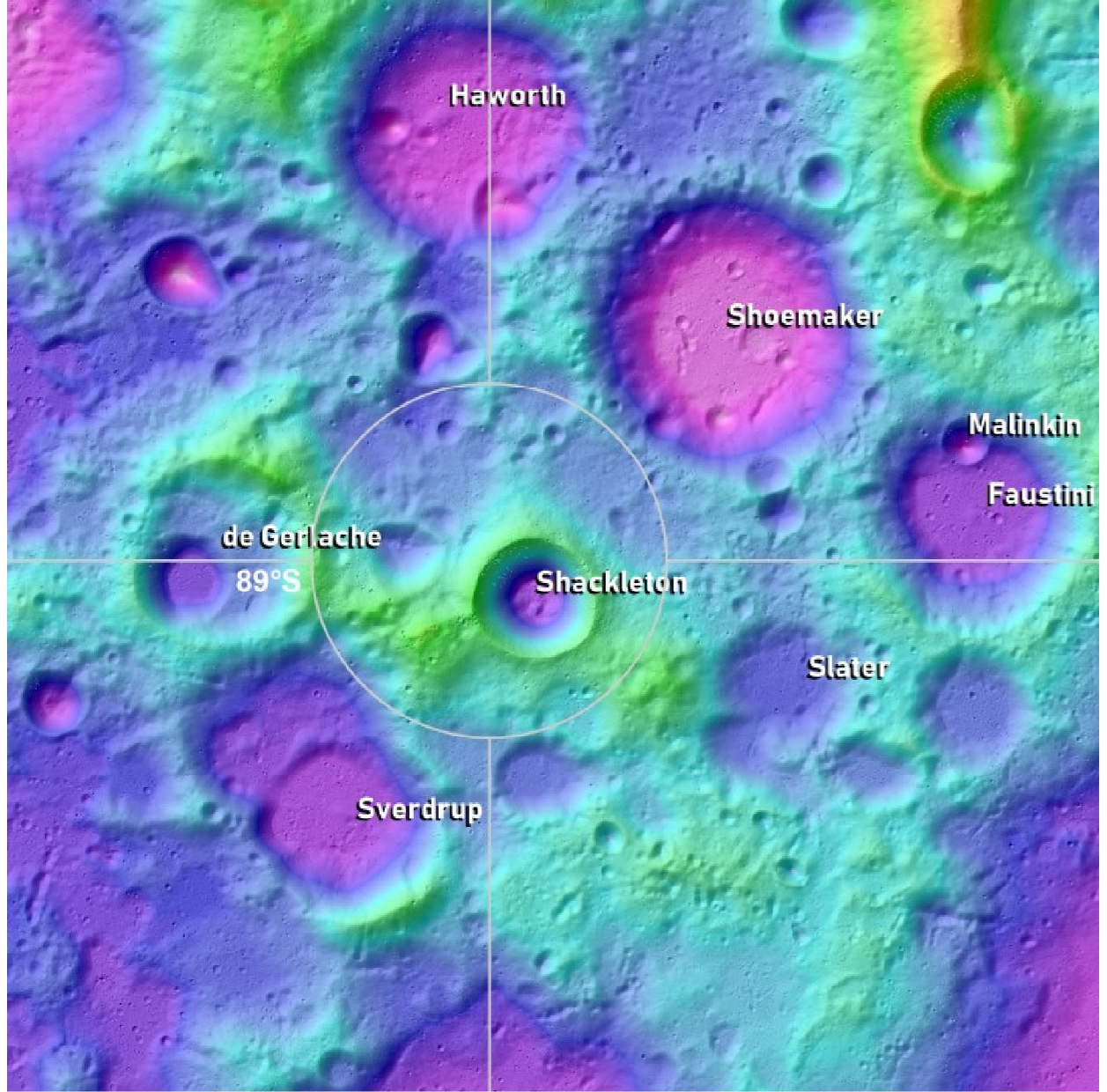}
        \caption{}
        \label{fig:2a}
    \end{subfigure}
    \hspace{0.03\textwidth}
    \begin{subfigure}{0.4\textwidth}
        \includegraphics[height=1.8in]{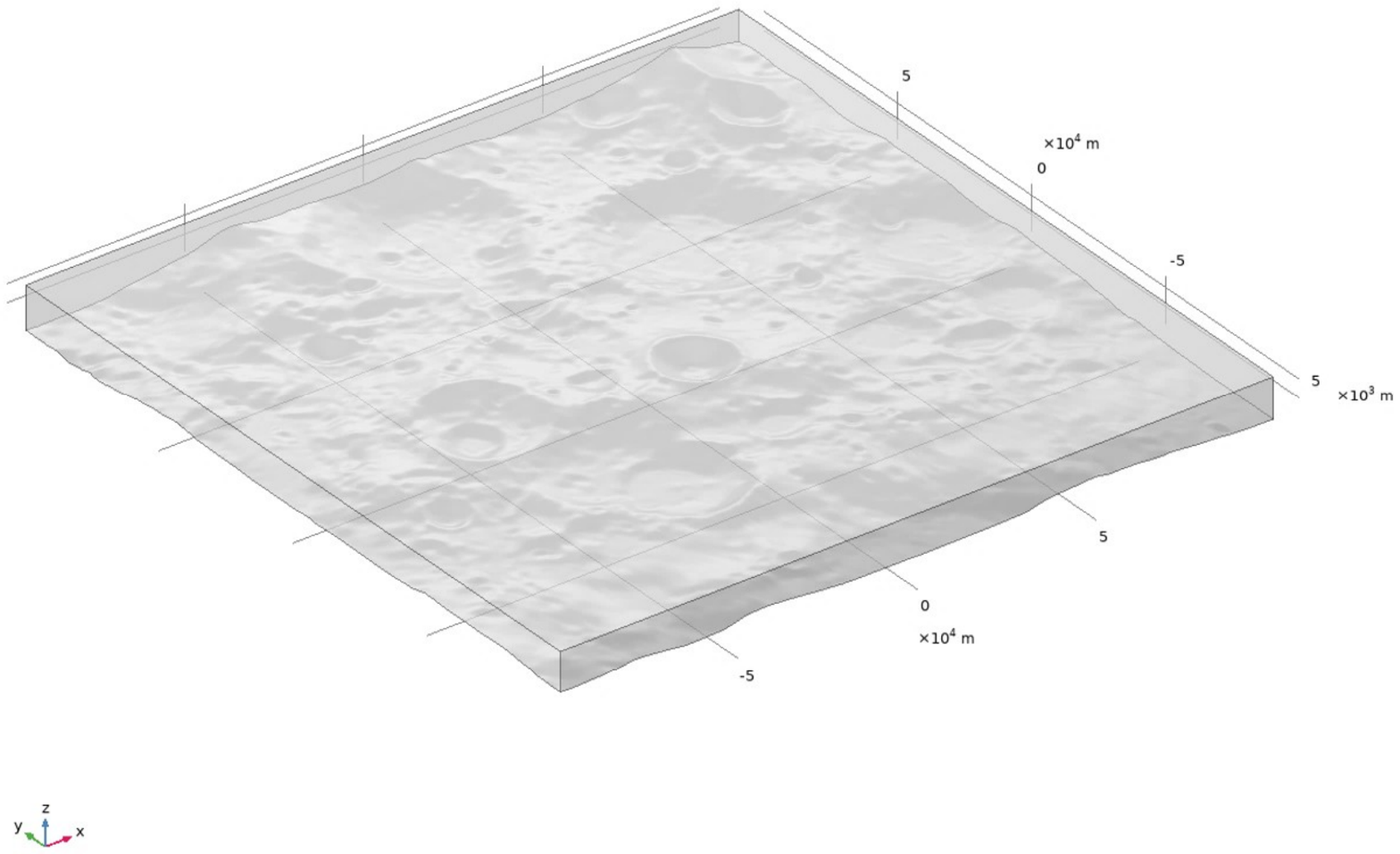}
        \caption{}
        \label{fig:2b}
    \end{subfigure}
    \caption{(\subref{fig:2a}) The topographic map of the lunar south pole in top view. (\subref{fig:2b}) The geometric model.}
    \label{fig:2}
\end{figure}

This work primarily focuses on the area between 86°S and 90°S. Centered at the lunar south pole, we extracted topographic data within a square region with a side length of 200 km to construct the geometric model of the lunar south pole surface. The computational domain extends from the lunar surface up to an altitude of 8000 m. Sunlight and plasma enter the computational domain in parallel to the xOz plane from one lateral boundary, with the incident angle relative to the x-axis defined as the solar elevation angle. Figure 2(\subref{fig:2a}) presents the topographic map of the lunar south pole in top view\cite{Stopar1}, and our geometric model is shown in Figure 2(\subref{fig:2b}). This model covers six major craters near the south pole, namely Shackleton, Haworth, Shoemaker, Faustini, de Gerlache and Sverdrup.

\subsection{Coupled Simulation Model}
In this study, the finite element method(FEM) is employed to calculate the electrostatic field, plasma environment and illumination conditions within the computational domain. Most previous studies adopted numerical methods to compute surface potential, which suffer from low computational efficiency. This limitation becomes particularly prominent when simulating half of the lunar orbital period in this study. The use of a neural network can greatly improve computational efficiency\cite{ZhangH1,LiuH1}. Therefore, a back propagation (BP) neural network is utilized in the present work for potential prediction. Following the work of Song et al.\cite{song1}, the input variables of the neural network include electron density, ion density, electron temperature, ion temperature, solar radiation, work function, maximum secondary electron yield, energy for maximum secondary electron yield, relative permittivity and conductivity. The output variable is the surface charging potential. The dataset used for network training covers plasma densities ranging from $9 \times 10^4$ to $1 \times 10^9\ \mathrm{m}^{-3}$ and plasma temperatures from 3 eV to 2000 eV, encompassing all plasma environmental conditions encountered during the Moon's transit through the Earth's magnetosphere. The output potential is calculated based on the current balance equation\cite{pandya1,wang2}:
\begin{equation} \label{eq:1}
C \frac{dU}{dt} = -J_e + J_i + J_{se} + J_{si} + J_{be} + J_{ph} - J_c
\end{equation}
where $J_e$ is the electron current density, $J_i$ is the ion current density, $J_{se}$ is the secondary electron current density induced by electrons, $J_{si}$ is the secondary electron current density induced by ions, $J_{be}$ is the backscattered electron current density, $J_{ph}$ is the photoelectron current density, and $J_c$ is the conduction current density. $U$ represents the potential of the lunar surface, and $C$ is the capacitance per unit area for a certain thickness of the lunar surface layer.

The surface potential is obtained by feeding environmental conditions into the trained neural network. Subsequently, the potential and electric field distribution across the computational domain are solved using the Poisson equations for the electrostatic field:
\begin{equation}
\begin{cases}
\nabla \cdot \mathbf{D} = \rho_v \\
\mathbf{D} = \varepsilon_0 \varepsilon_r \mathbf{E} \\
\mathbf{E} = -\nabla U
\end{cases}
\label{eq:maxwell_electric}
\end{equation}

Electrons have a small mass, high thermal velocity and high mobility, so their spatial distribution is highly sensitive to local potential variations. Accordingly, both density-gradient-driven diffusion and electric-field-driven drift are considered for electron transport, which is calculated using the drift-diffusion equation:
\begin{equation}
\begin{cases}
\dfrac{\partial n_e}{\partial t} + \nabla\cdot \boldsymbol{\Gamma}_e = R_e - \left(\mathbf{u}\cdot\nabla\right) n_e \\[4pt]
\boldsymbol{\Gamma}_e = -\mu_e \mathbf{E} n_e - D_e \nabla n_e \\[4pt]
\dfrac{\partial n_{\varepsilon e}}{\partial t} + \nabla\cdot \boldsymbol{\Gamma}_{\varepsilon e} + \mathbf{E}\cdot\boldsymbol{\Gamma}_e = S_{en} - \left(\mathbf{u}\cdot\nabla\right) n_{\varepsilon e} + \dfrac{Q+Q_{\mathrm{gen}}}{e} \\[4pt]
\boldsymbol{\Gamma}_{\varepsilon e} = -\mu_{\varepsilon e} \mathbf{E} n_{\varepsilon e} - D_{\varepsilon e} \nabla n_{\varepsilon e}
\end{cases}
\label{eq:transport_system}
\end{equation}
where $n_e$ denotes the electron density, $n_{\varepsilon e}$ is the average electron energy, and $\mathbf{u}$ is the velocity vector.

The mobility and diffusion coefficient are expressed as follows:
\begin{equation}
\begin{cases}
\mu_e = \dfrac{e}{m_e \nu_c} \\[6pt]
D_e = \dfrac{k T_e}{m_e \nu_c}
\end{cases}
\label{eq:mob_diff}
\end{equation}

Compared with electrons, ions have larger mass and lower mobility. Their motion is predominantly governed by the overall movement of the solar wind. Accordingly, the advection-diffusion equation is adopted to describe ion transport:
\begin{equation} \label{eq:ion_transport}
\dfrac{\partial c_i}{\partial t} + \nabla\cdot \mathbf{J}_i + \mathbf{u}\cdot\nabla c_i = R_i
\end{equation}
where $c_i$ denotes the ion concentration, $R_i$ represents the ion reaction rate term, and $\mathbf{J}_i$ stands for the diffusion flux vector. The expression for $\mathbf{J}_i$ is given below:
\begin{equation} \label{eq:Ji_def}
\mathbf{J}_i = -D_i \nabla c_i - z_i u_{m,i} F c_i \nabla V
\end{equation}
where $z_i$ is the ion charge number, $F$ is the Faraday constant, $V$ is the electric potential, $u_{m,i}$ is the ion mobility, and $D_i$ is the diffusion coefficient. The corresponding expressions are given as follows:
\begin{equation} \label{eq:Di}
D_i = \dfrac{k T_i}{m_i \nu_c}
\end{equation}

Under complex topographic conditions, solar radiation intensity depends not only on the solar elevation angle but also on terrain shading. In this study, the Transport of Diluted Species (tds) interface is adopted to simulate the shadowing effect induced by topography. The solar radiation intensity over the entire surface is obtained by multiplying the shading factor by the cosine of the angle between the incident direction of sunlight and the surface normal vector for correction.

The physical model for lunar surface charging, the finite element–neural network coupled computational workflow, and the BP neural network-based potential prediction method used in this study are derived from a previously validated modeling framework\cite{song1}. By comparison with conventional numerical calculations, this framework has been shown to effectively predict lunar surface charging potentials under various plasma environments and material parameter conditions.

\subsection{Simulation Parameter Settings}

This study investigates the effects of varying solar elevation angles and magnetospheric plasma conditions on lunar surface charging. For the analysis of solar elevation angle, the plasma environment is set to the solar wind regime, with a plasma density of $5 \times 10^6\ \mathrm{m}^{-3}$, electron temperature of 10 eV, ion temperature of 12 eV, and solar wind velocity of 400 km/s. Given that the solar elevation angle at the lunar south pole remains extremely low and ranges from 0° to 2°\cite{derosa1}, only three values (0°, 1° and 2°) are adopted in this work.

\begin{figure}[h]
    \centering
    \includegraphics[width=0.5\textwidth]{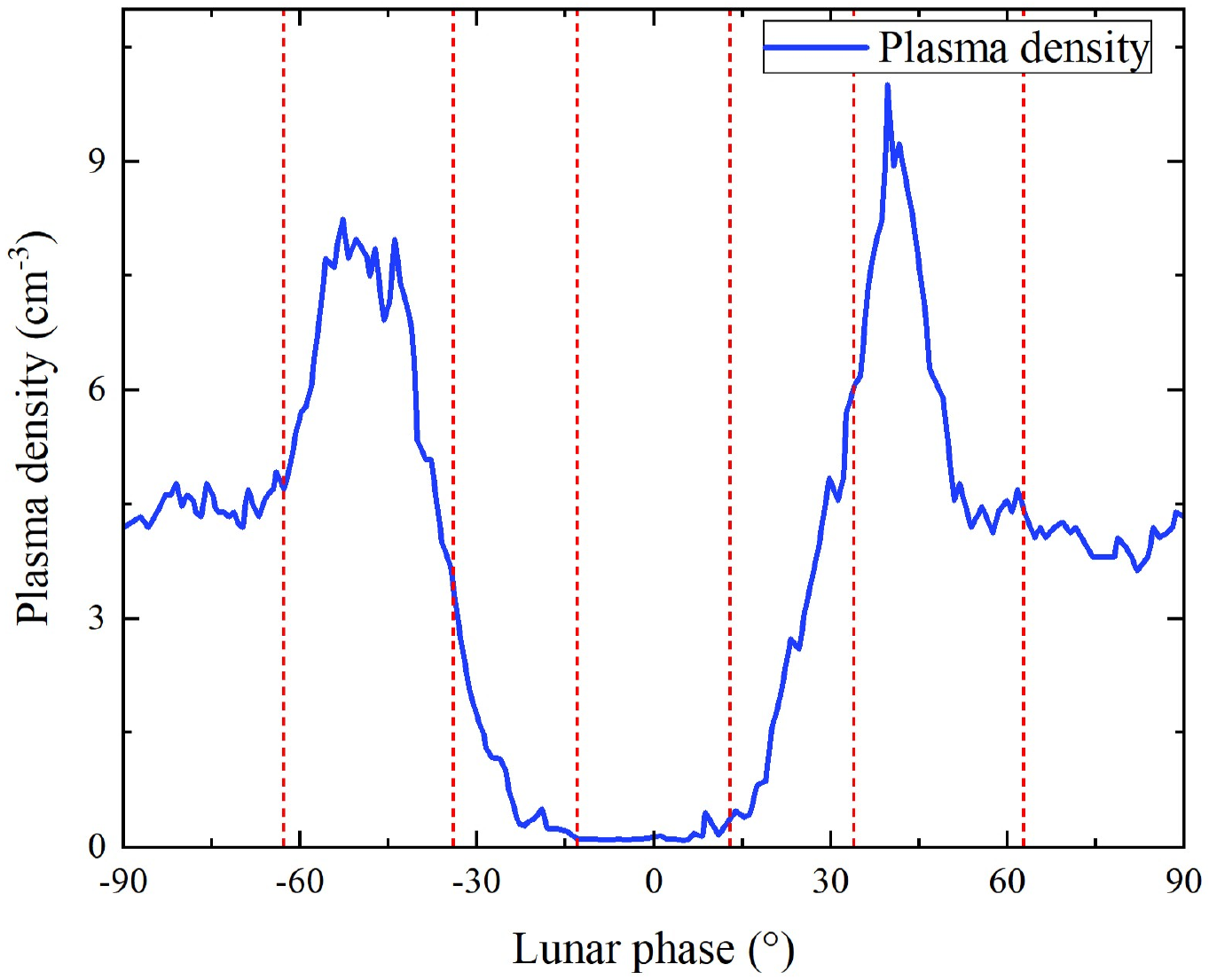}
    \caption{Curve of plasma density versus lunar phase.}
    \label{fig:3}
\end{figure}

When investigating the effects of different plasma environments, plasma density and temperature are set as time-dependent inputs. The total simulation duration is half of the lunar orbital period, namely 1180224 seconds. This simulation covers the entire process in which the Moon travels from the solar wind into the Earth's magnetosphere and passes through the magnetosheath, magnetotail lobe and plasma sheet. The plasma density data are derived from measurements acquired by the ARTEMIS mission\cite{LiS1}. The temporal variation of plasma density is plotted in Figure \ref{fig:3}. Several discrete points are extracted from the curve to construct an interpolation function, which is invoked in the model to obtain density values at arbitrary moments. The red dashed lines in the figure mark the boundaries between the solar wind environment and the different regions of the Earth’s magnetosphere, corresponding to x-coordinates of −62.7°, −34°, −12.9°, 12.9°, 34° and 62.7°. The regions are symmetric about the lunar phase of 0°, arranged from outside to inside as the solar wind, magnetosheath, magnetotail lobe, and plasma sheet. The red dashed lines in subsequent figures have the same meaning.

\begin{figure}[h]
    \centering
    \begin{subfigure}{0.4\textwidth}
        \includegraphics[height=2in]{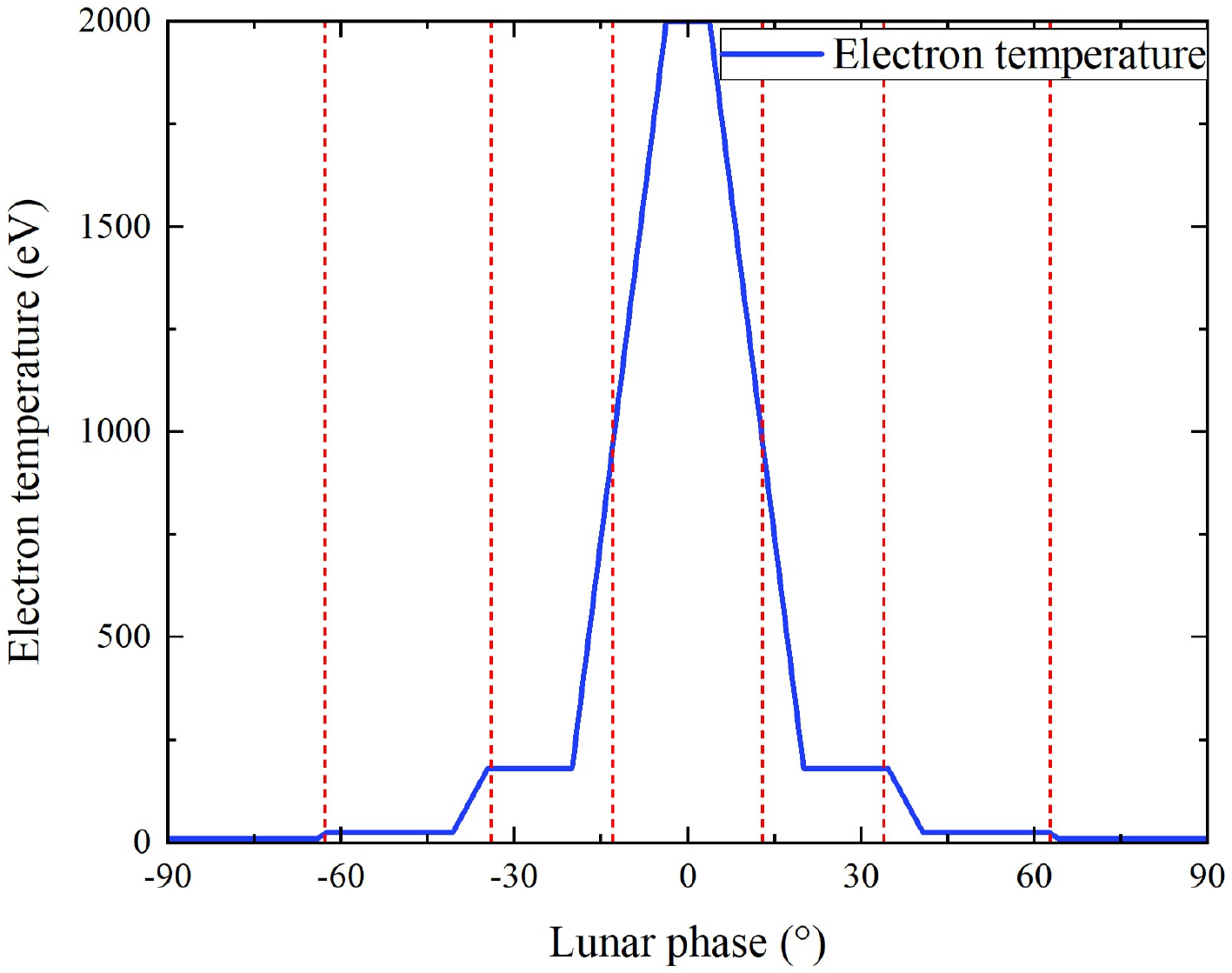}
        \caption{}
        \label{fig:4a}
    \end{subfigure}
    \hspace{0.03\textwidth}
    \begin{subfigure}{0.4\textwidth}
        \includegraphics[height=2in]{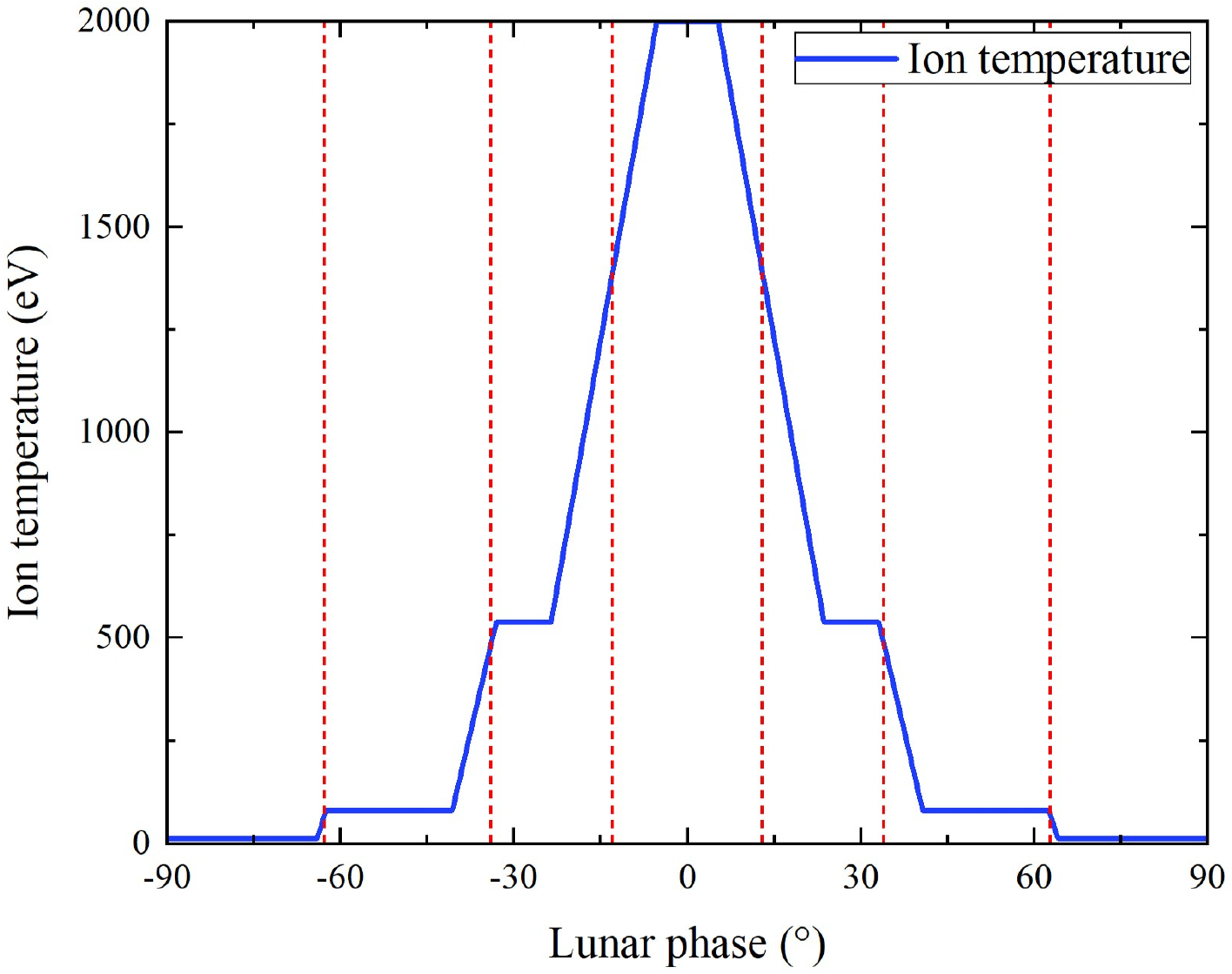}
        \caption{}
        \label{fig:4b}
    \end{subfigure}
    \caption{(\subref{fig:4a}) Curve of electron temperature versus lunar phase. (\subref{fig:4b}) Curve of ion temperature versus lunar phase.}
    \label{fig:4}
\end{figure}

The electron and ion temperatures of different plasma environments are listed in Table \ref{tab:1}\cite{zeng1}. Since abrupt changes in plasma temperature would result in numerical non-convergence during computation, transition intervals are introduced between adjacent time segments. Within these intervals, the temperature varies linearly with time instead of stepping instantaneously. The temporal profile of plasma temperature is presented in Figure \ref{fig:4}.
\begin{table}[h]
    \caption{Plasma temperatures of different plasma environments.}    
    \centering
    \begin{tabular}{ccc}
            \hline
            Plasma environment & Electron temperature(eV) & Ion temperature(eV) \\  
            \hline
            solar wind & 10 & 12\\ 
            magnetosheath & 26 & 80 \\
            magnetotail lobe & 180 & 540 \\
            plasma sheet & 2000 & 2000 \\
            \hline
            \end{tabular}

    \label{tab:1}
\end{table}

Uniform material parameters are adopted across all models. Specifically, the work function is 5.58 eV, the maximum secondary electron yield equals 1, the energy for maximum secondary electron yield is 350 eV, the relative permittivity is 5.5, and the conductivity is 7 pS/m.

\section{Results and Discussion}
\subsection{Influence of Topography}
\begin{figure}[h]
    \centering
    \begin{subfigure}{0.4\textwidth}
        \includegraphics[height=2in]{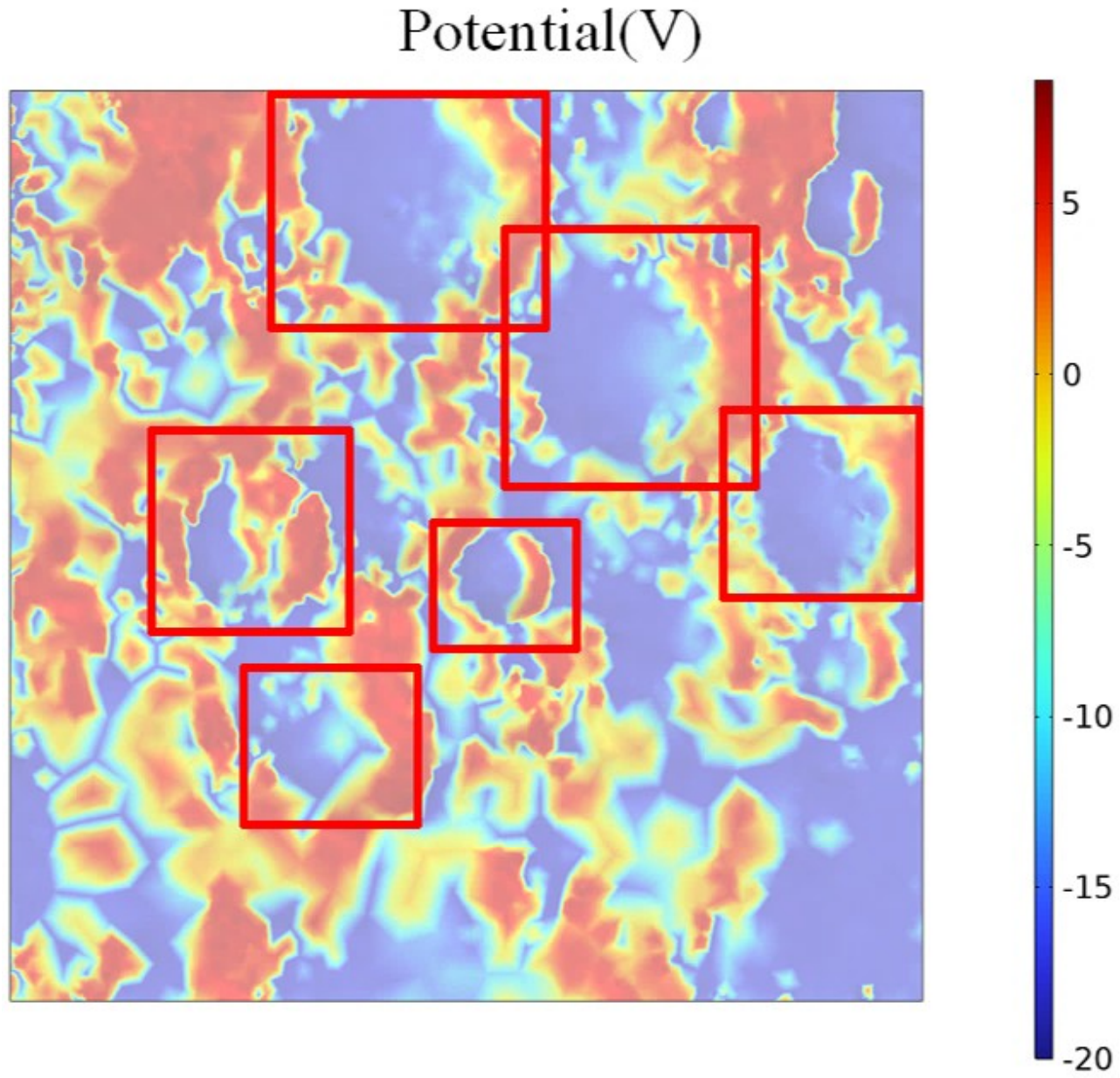}
        \caption{}
        \label{fig:5a}
    \end{subfigure}
    \hspace{0.03\textwidth}
    \begin{subfigure}{0.4\textwidth}
        \includegraphics[height=2in]{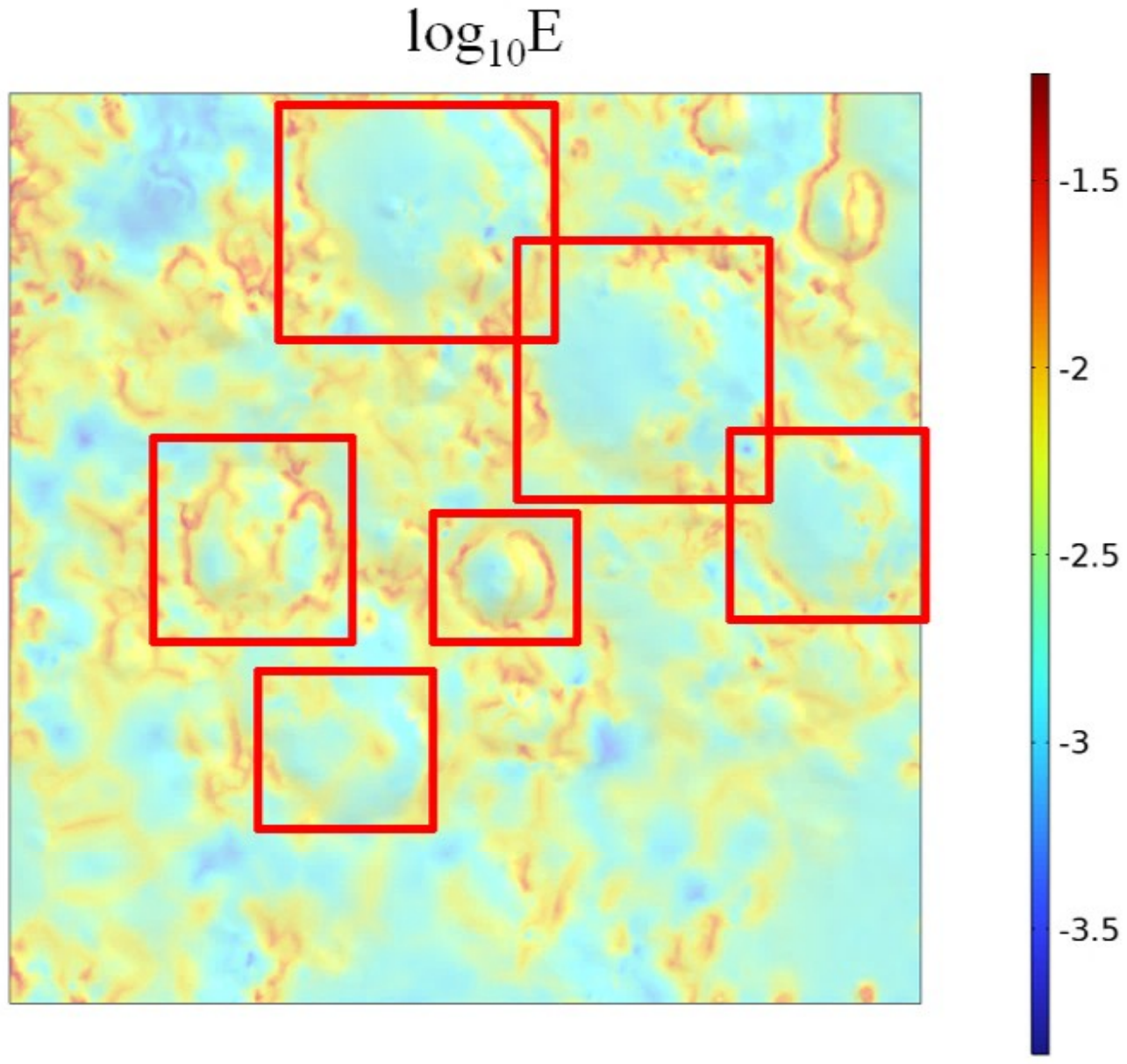}
        \caption{}
        \label{fig:5b}
    \end{subfigure}
    \caption{(\subref{fig:5a}) Distribution of surface potential (V) over the lunar south pole at a solar elevation angle of 1° under solar wind conditions. (\subref{fig:5b}) Distribution of surface electric field magnitude (V/m) over the lunar south pole at a solar elevation angle of 1° under solar wind conditions, plotted on a logarithmic scale.}
    \label{fig:5}
\end{figure}

Figure \ref{fig:5} presents the surface potential distribution and electric field intensity distribution across the lunar south polar region under solar wind conditions at a solar elevation angle of 1°.

\begin{figure}[h]
    \centering
    \begin{subfigure}{0.4\textwidth}
        \includegraphics[height=2in]{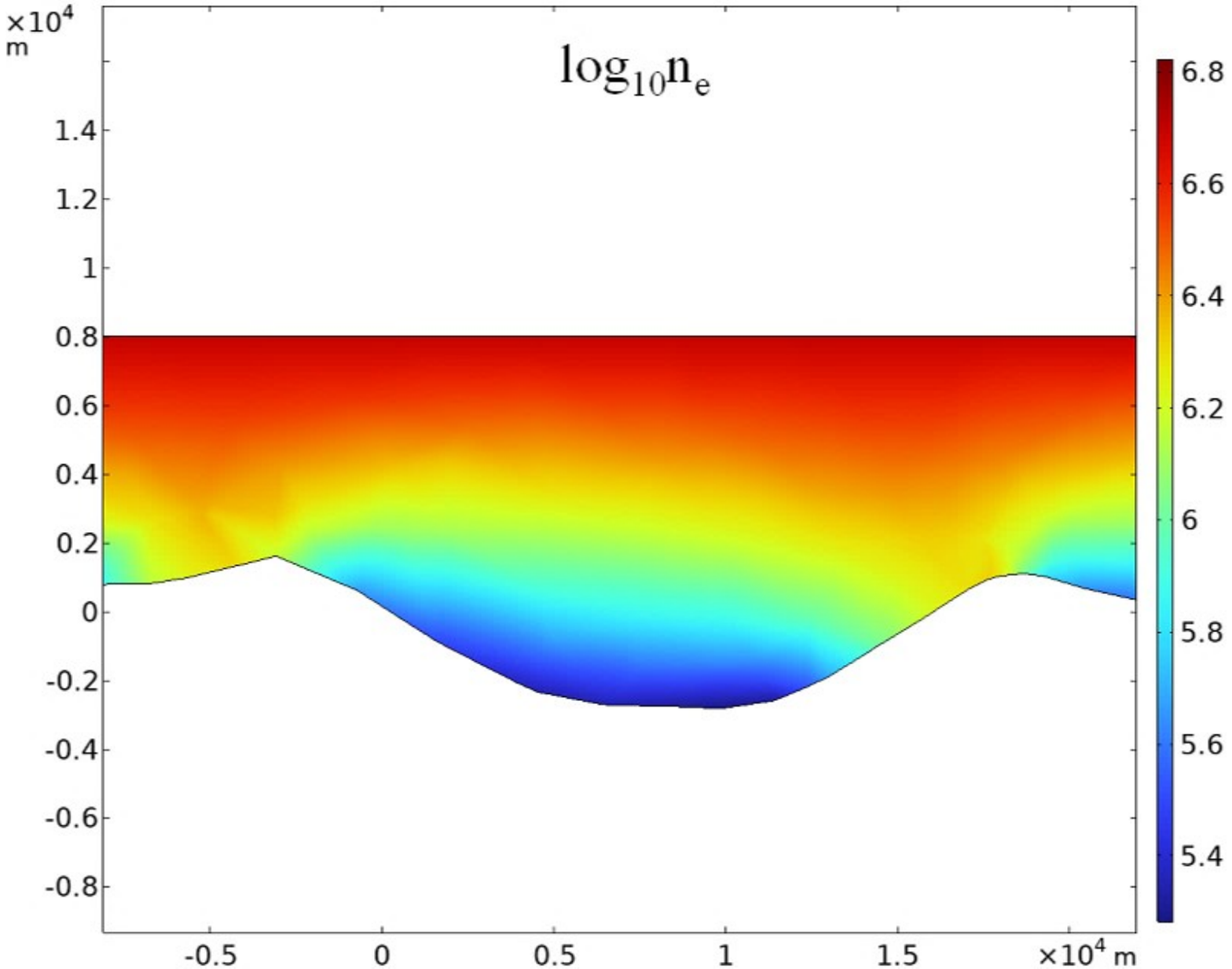}
        \caption{}
        \label{fig:6a}
    \end{subfigure}
    \hspace{0.03\textwidth}
    \begin{subfigure}{0.4\textwidth}
        \includegraphics[height=2in]{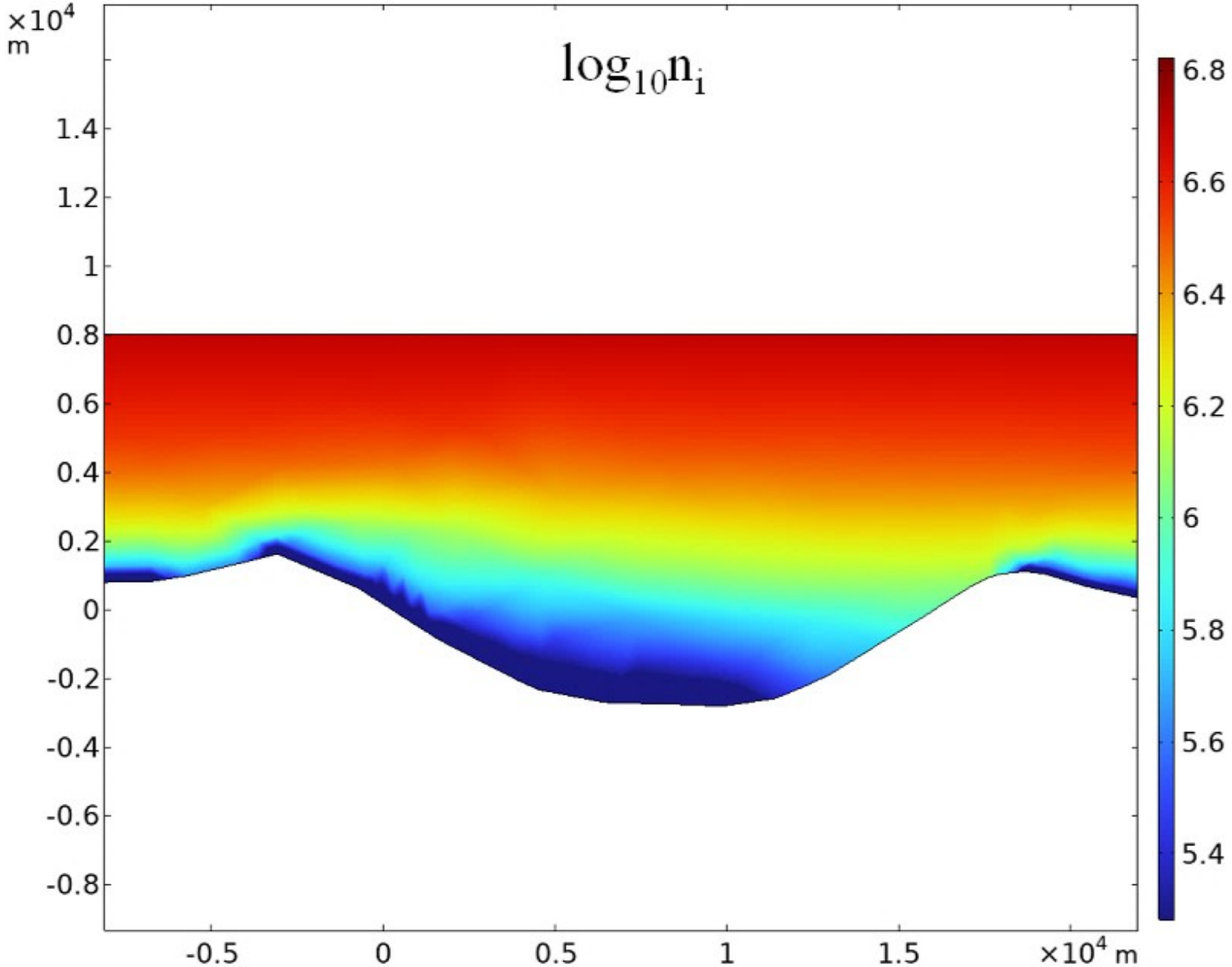}
        \caption{}
        \label{fig:6b}
    \end{subfigure}

    \vspace{0.03\textwidth}

    \begin{subfigure}{0.4\textwidth}
        \includegraphics[height=2in]{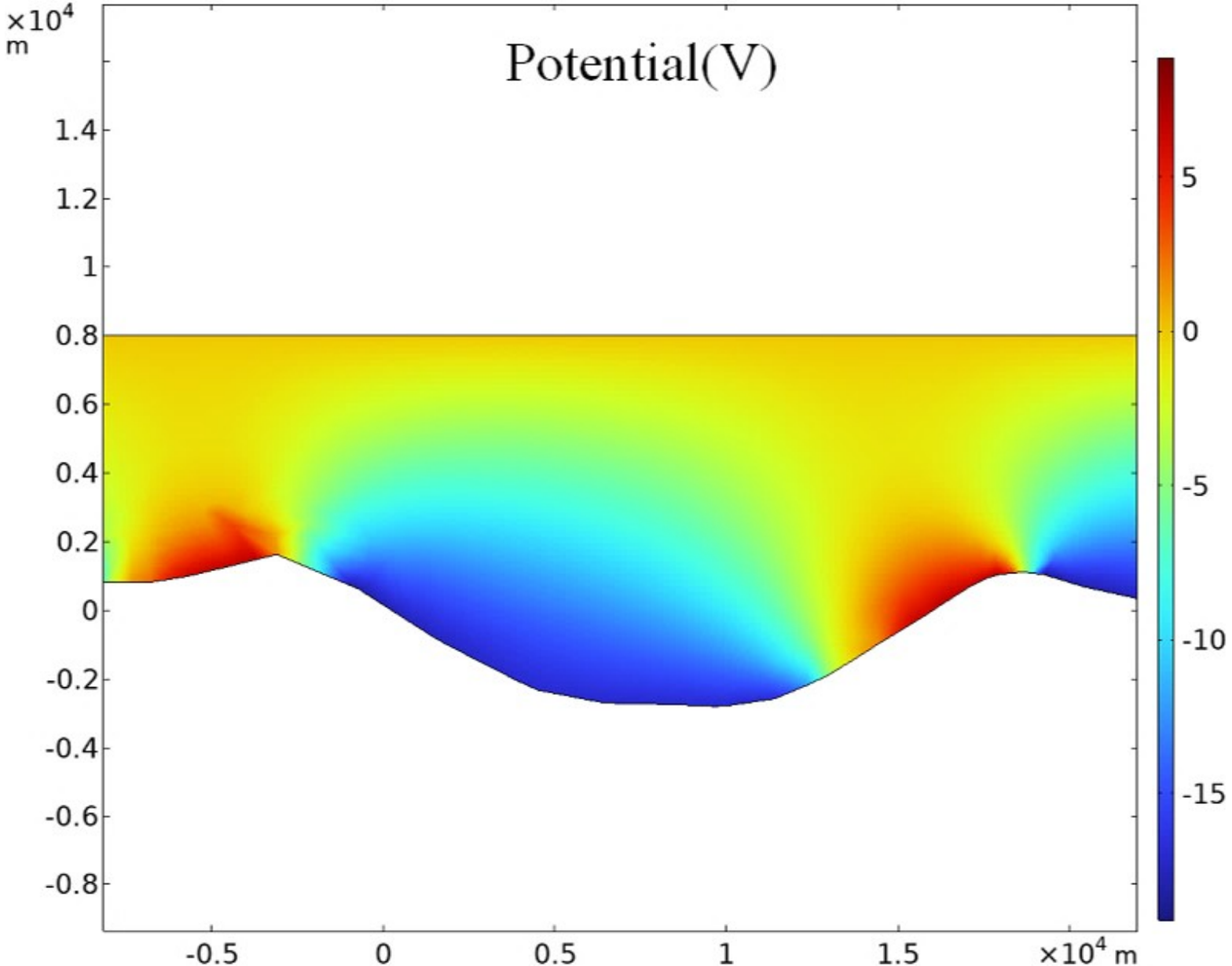}  
        \caption{}
        \label{fig:6c}
    \end{subfigure}
    \hspace{0.03\textwidth}
    \begin{subfigure}{0.4\textwidth}
        \includegraphics[height=2in]{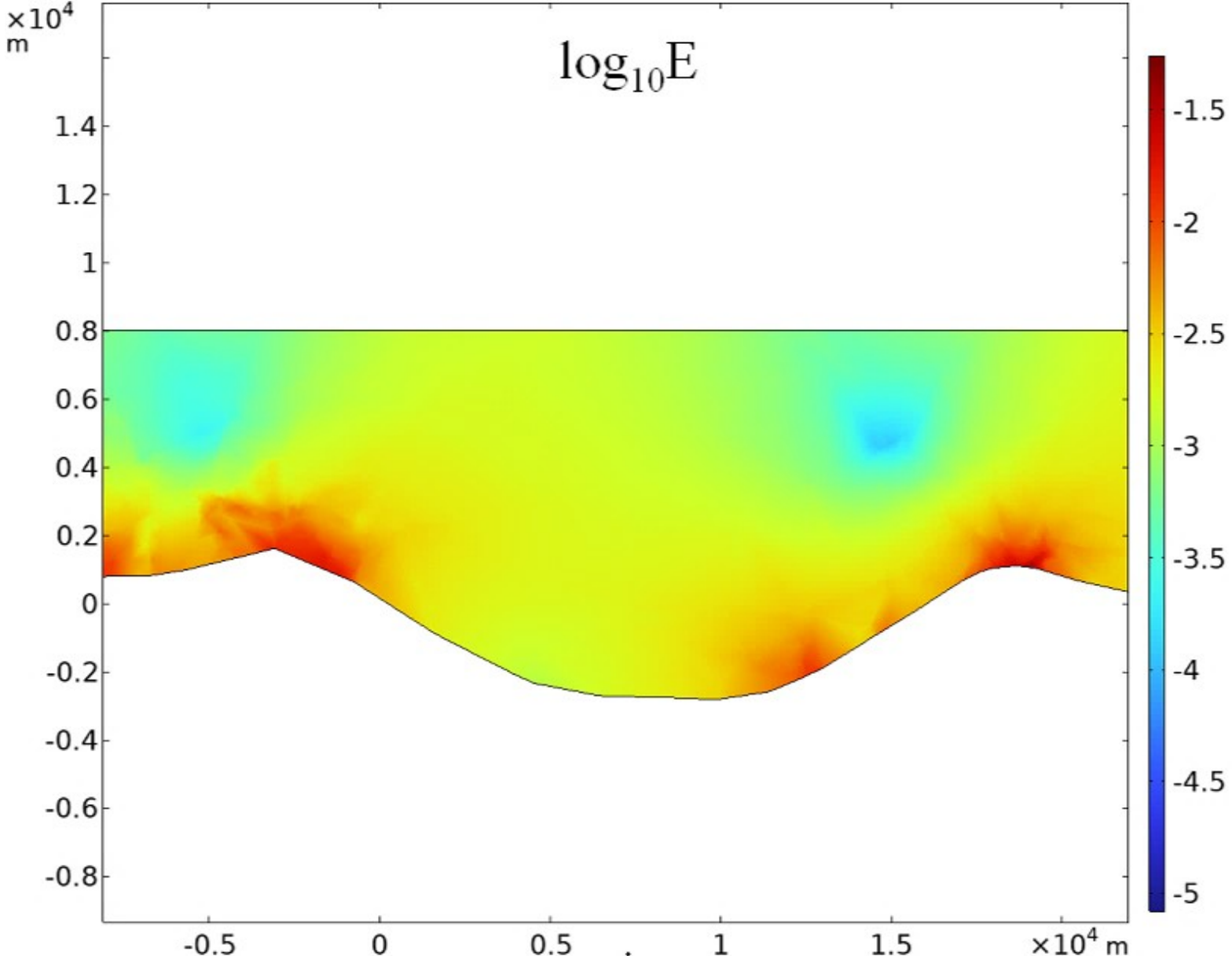} 
        \caption{}
        \label{fig:6d}
    \end{subfigure}

    \caption{Distributions of electron density, ion density, potential and electric field magnitude along the cross-section of Shackleton Crater at a solar elevation angle of 1° under solar wind conditions. Electron density, ion density and electric field magnitude are plotted on a logarithmic scale.
    (\subref{fig:6a}) Electron density ($\mathrm{m^{-3}}$) (\subref{fig:6b}) Ion density ($\mathrm{m^{-3}}$) (\subref{fig:6c}) Potential (V). (\subref{fig:6d}) Electric field magnitude (V/m).}
    \label{fig:6}
\end{figure}

Overall, the surface potential within the study area exhibits a distinct topographic dependence. Relatively high potentials tend to form on the windward sides of raised landforms, whereas considerably lower potentials are observed in topographically shadowed regions such as leeward sides and crater floors. This distribution pattern indicates that under extremely low solar elevation angles, the complex topography of the lunar south pole modulates both solar illumination and the incidence of solar wind plasma, thereby inducing prominent charging discrepancies among different geomorphic units. The potential distribution along the cross-section of Shackleton Crater in Figure \ref{fig:6} further illustrates this characteristic, where the crater walls represent typical raised landforms.

From a physical perspective, the lunar surface potential is not directly governed by topography, but rather determined by local current balance. At any given surface location, the equilibrium potential is collectively controlled by incident electron current, incident ion current, photoelectron emission current and secondary electron emission current. For illuminated surfaces, enhanced photoelectron emission current partially counteracts the solar wind electron current, driving the surface potential toward positive values. In contrast, in shadowed regions, photoelectron emission is drastically suppressed, and electron current becomes dominant. Consequently, the surface develops a more negative potential to suppress incoming electrons and facilitate ion collection, until a new current equilibrium is established. 

At a solar elevation angle of 1°, sunlight propagates almost tangentially along the lunar surface. Crater rims, local highlands and small-scale topographic undulations significantly amplify the shadowing effect, leaving leeward areas deprived of direct solar irradiation and weakening photoelectron emission. By comparison, windward slopes are fully exposed to direct sunlight; meanwhile, the large angle between the solar incident direction and steep slopes leads to relatively intense solar radiation on windward sides. Beyond shadowing solar irradiation, raised landforms also obstruct plasma flow. As revealed by the electron and ion density distributions in Figure \ref{fig:6}, plasma is blocked by crater walls, giving rise to plasma cavities on the leeward side. Within such cavities, ion density is lower than electron density. Ions possess larger mass and greater inertia, so their trajectories are predominantly governed by the bulk flow direction of solar wind. In comparison, electrons possess a higher thermal velocity and superior thermal diffusivity, allowing them to migrate into the cavity from the surrounding regions more rapidly. This enhances the contribution of electron current to surface charging in cavity regions, resulting in much lower surface potentials.

The electric field distributions in Figure \ref{fig:5} and Figure \ref{fig:6} show that strong electric fields generally emerge near raised landforms and the junction between crater floors and downstream crater walls. This is attributed to the sharp variations in surface potential across these zones, which give rise to large local potential gradients.The maximum magnitude of electric field intensity throughout the computational domain is approximately 0.1 V/m, which is a relatively low value. This is closely associated with the model scale. As this study focuses on kilometer-scale terrain consisting of multiple large craters in the lunar south pole, surface potential differences generally vary gradually over spatial scales ranging from hundreds to thousands of meters. Even with a local potential difference of more than ten volts, the large-scale volume electric field remains on the order of $10^{-1}$ V/m.

\begin{table}[h]
    \caption{The depth-to-width ratios and floor potentials of six major craters at a solar elevation angle of 1° under solar wind conditions.}    
    \centering
    \begin{tabular}{ccc}
            \hline
            Crater & Depth-to-width ratio & Floor potential(V) \\  
            \hline
            Shackleton & 0.125 & -17.079\\ 
            Haworth & 0.067 & -16.847 \\
            Shoemaker & 0.066 & -16.798 \\
            Faustini & 0.076 & -16.761 \\
            De gerlache & 0.091 & -17.199 \\
            Sverdrup & 0.091 & -17.064 \\
            \hline
            \end{tabular}

    \label{tab:2}
\end{table}

Table \ref{tab:2} lists the depth-to-width ratios and floor potentials of six major craters. The floor potential of each crater is approximately −17 V. The depth-to-width ratio exerts a certain influence on crater floor potential: craters with larger depth-to-width ratios usually produce stronger shadowing for both solar illumination and plasma incidence, which tends to yield more negative floor potentials. Nevertheless, since the six selected craters generally have small depth-to-width ratios, and crater floor potential is jointly affected by additional factors including wall slope, rim height and shielding from adjacent terrain, the relationship between depth-to-width ratio and floor potential does not follow a strict monotonic trend.

\subsection{Influence of Solar Elevation Angle}

\begin{figure}[h]
    \centering
    \begin{subfigure}{0.3\textwidth}
        \includegraphics[height=1.5in]{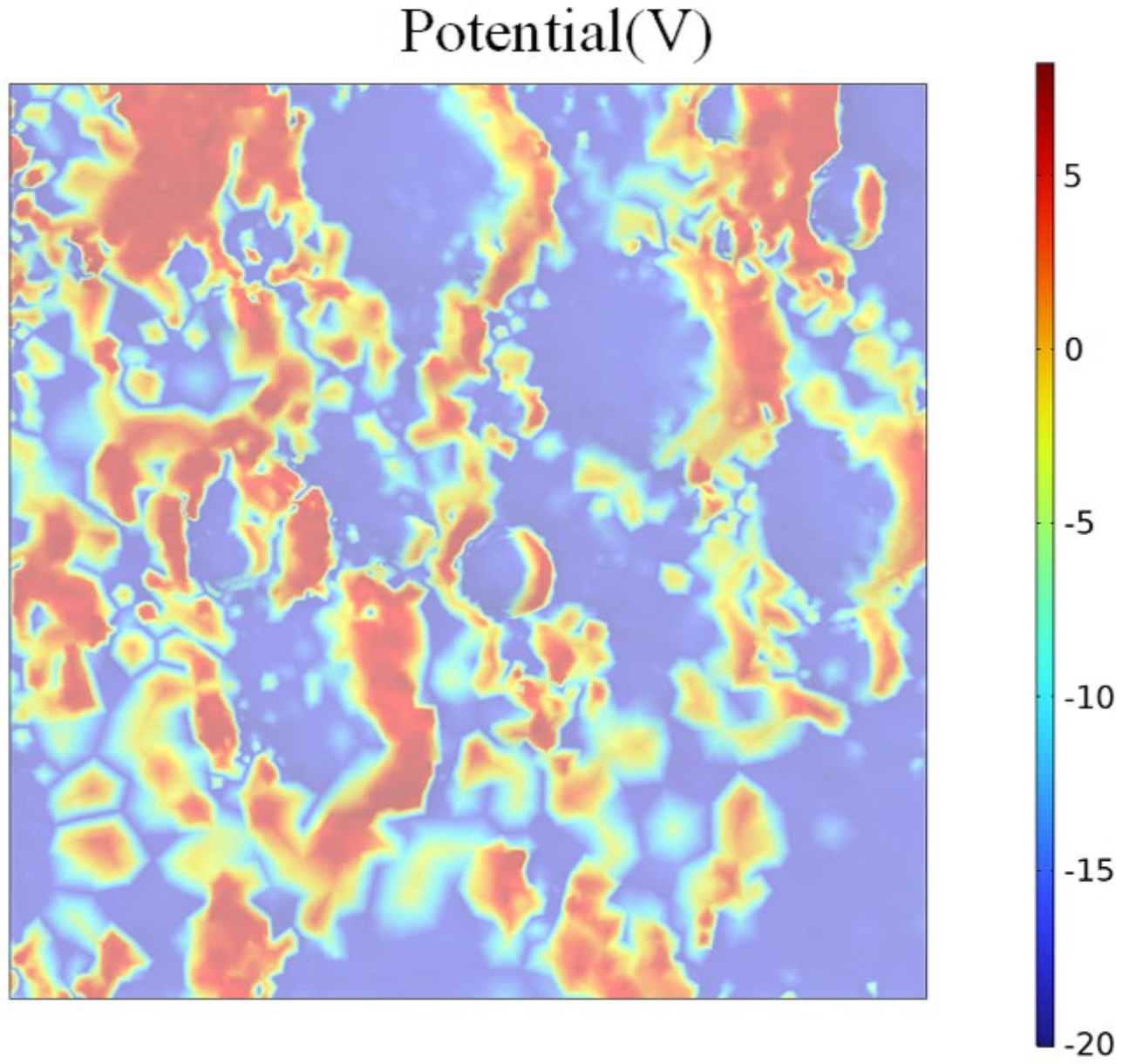}
        \caption{}
        \label{fig:7a}
    \end{subfigure}
    \begin{subfigure}{0.3\textwidth}
        \includegraphics[height=1.5in]{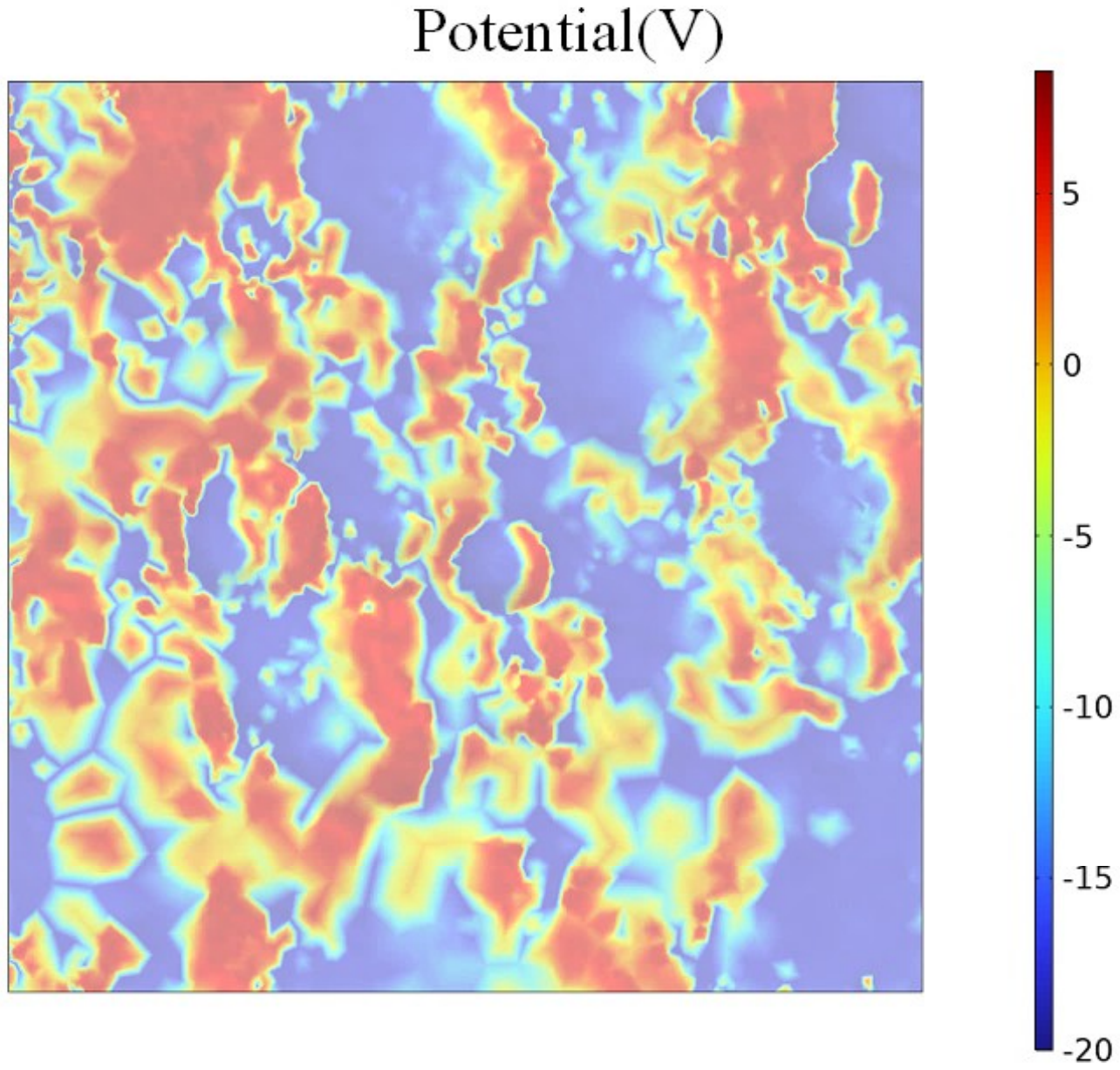}
        \caption{}
        \label{fig:7b}
    \end{subfigure}
    \begin{subfigure}{0.3\textwidth}
        \includegraphics[height=1.5in]{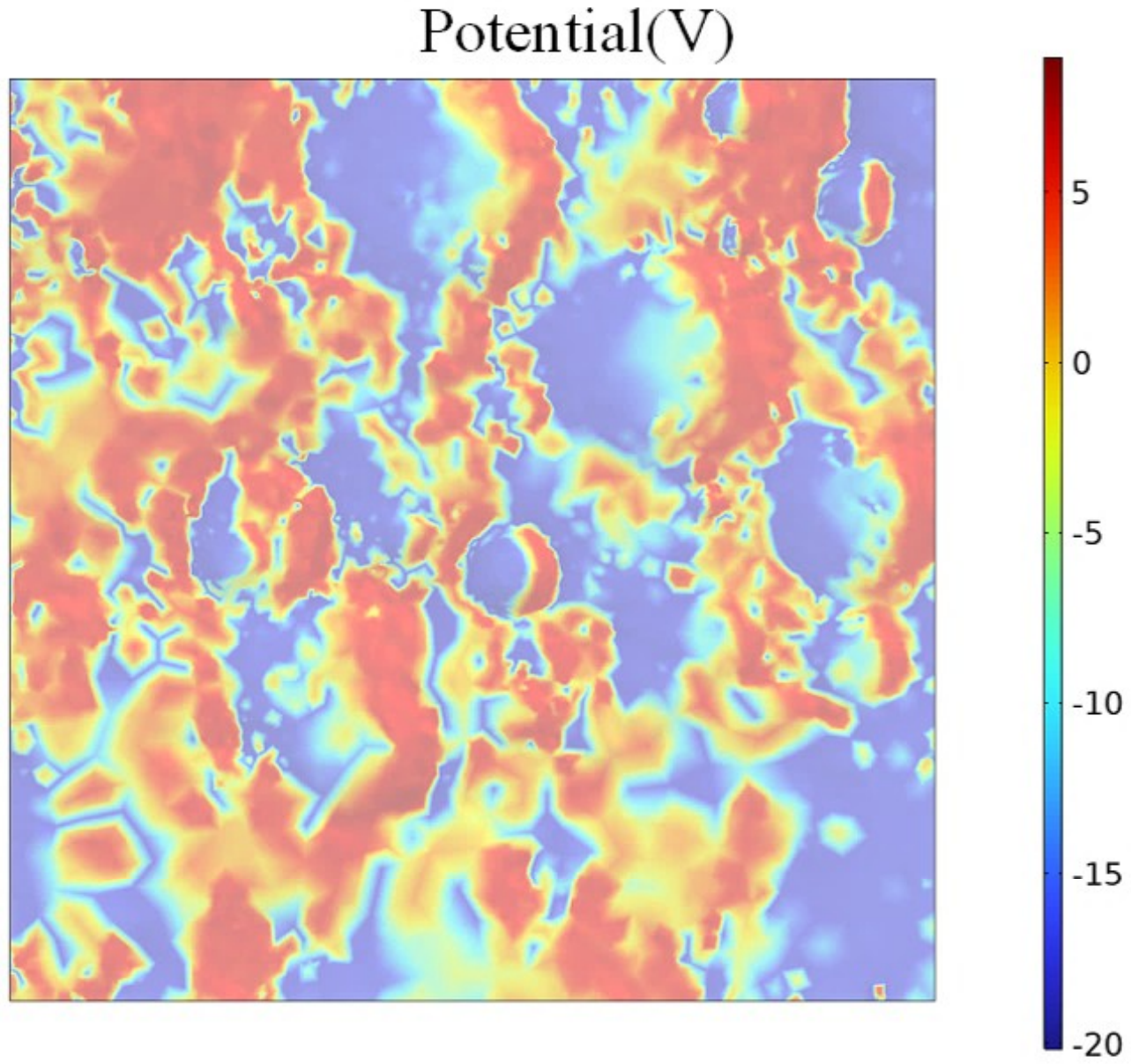}
        \caption{}
        \label{fig:7c}
    \end{subfigure}
    \caption{Surface potential (V) distributions at different solar elevation angles under solar wind conditions.
    (\subref{fig:7a}) Solar elevation angle of $0^\circ$. (\subref{fig:7b}) Solar elevation angle of $1^\circ$. (\subref{fig:7c}) Solar elevation angle of $2^\circ$.}
    \label{fig:7}
\end{figure}

Simulations were conducted to investigate surface charging at the lunar south pole under typical solar wind conditions with varying solar elevation angles. Figure \ref{fig:7} presents the surface potential distributions at solar elevation angles of 0°, 1° and 2°, respectively. Table \ref{tab:3} summarizes the maximum and minimum surface potentials corresponding to different solar elevation angles.

\begin{table}[h]
    \caption{The maximum and minimum surface potentials corresponding to different solar elevation angles.}    
    \centering
    \begin{tabular}{ccc}
            \hline
            Solar elevation angle($^\circ$) & Maximum surface potential(V) & Minimum surface potential(V) \\  
            \hline
            0 & 8.24 & -20.05\\ 
            1 & 8.6 & -20 \\
            2 & 8.75 & -19.47 \\
            \hline
            \end{tabular}

    \label{tab:3}
\end{table}

As illustrated in Figure \ref{fig:7}, the coverage of positive potential regions on the lunar surface expands evidently with the increase of solar elevation angle, and the boundary between positive and negative potentials inside craters gradually shifts toward the upstream crater wall. This phenomenon is primarily attributed to the weakened topographic shadowing effect. A larger solar elevation angle reduces the coverage of terrain shadow, and allows sunlight to reach the crater floors more easily. 

According to the data in Table \ref{tab:3}, the maximum surface potential rises continuously while the absolute value of negative potential decreases slightly as the solar elevation angle increases. Specifically, the maximum potential increases from 8.24 V to 8.75 V, with an increment of approximately 6.2\%; the absolute value of the minimum potential declines from 20.05 V to 19.47 V, representing a reduction of about 2.9\%. The increased solar elevation angle enhances the solar radiation on the lunar surface and intensifies photoelectron emission, thereby raising the surface potential of sunlit highland regions. In contrast, the negative potential at crater floors and permanently shadowed regions varies moderately due to the persistent limitation on photoelectron current. In addition, the variation of solar wind incident angle raises the probability of ions entering local cavities, which partially counteracts the negative potential induced by electron accumulation.

Further analysis on the potential responses of different geomorphic units reveals that the windward highlands are most sensitive to changes in solar elevation angle, with a relatively larger increase in surface potential. By comparison, crater floors, which are persistently located in shadowed areas and plasma shadow regions, show negligible responses to small variations in solar elevation angle, with nearly stable surface potentials. The results indicate that topographic shadowing dominates the surface charging process under extremely low solar elevation angles. Minor variations in solar elevation angle modulate local surface potentials mainly by strengthening photoelectron current. Consequently, windward highlands demonstrate the most pronounced response, whereas crater floors maintain relatively steady potentials.

\subsection{Influence of Plasma Environment}

\begin{figure}[!ht]
    \centering
    \begin{subfigure}{0.3\textwidth}
        \includegraphics[height=1.5in]{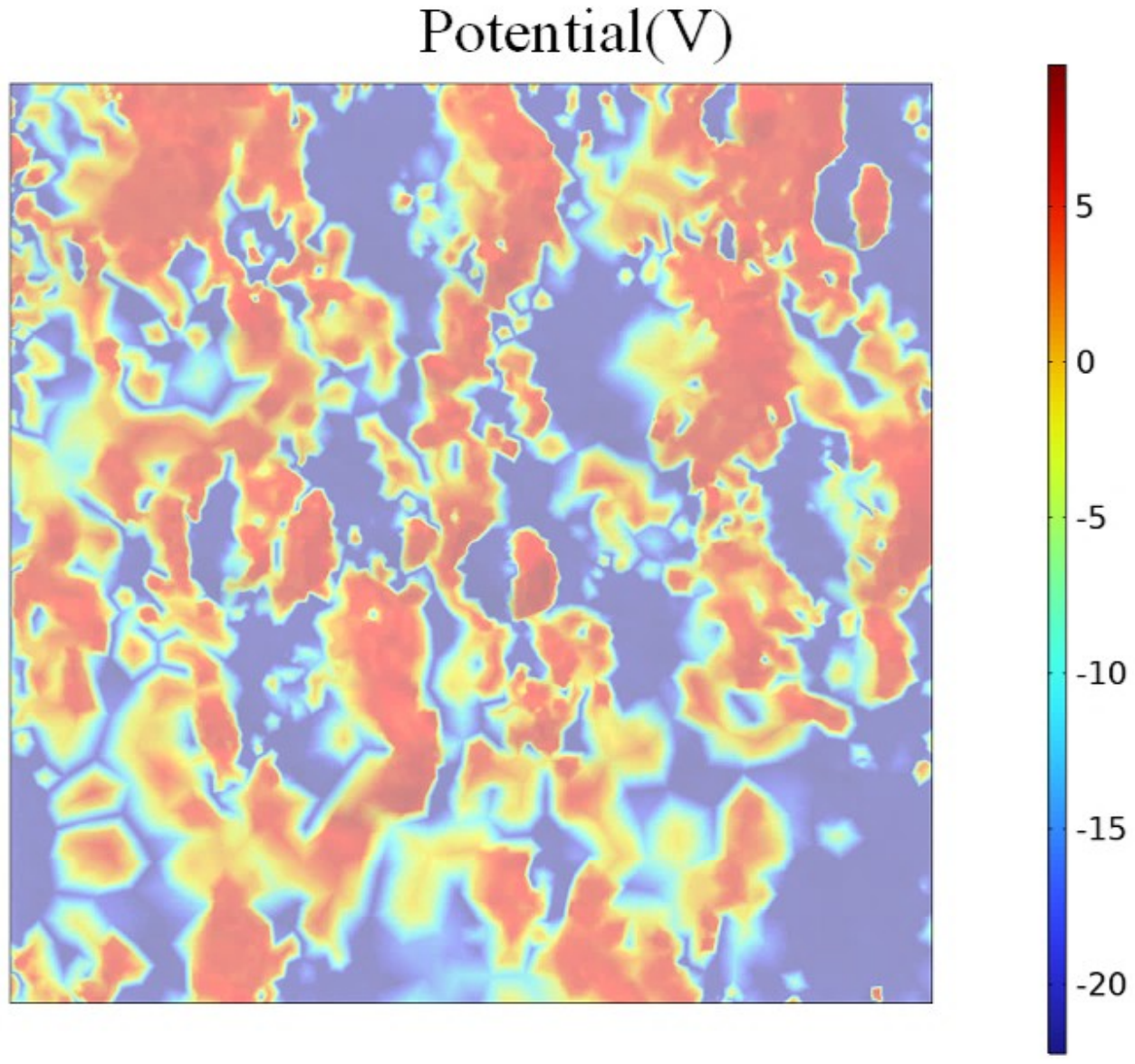}
        \caption{}
        \label{fig:8a}
    \end{subfigure}
    \begin{subfigure}{0.3\textwidth}
        \includegraphics[height=1.5in]{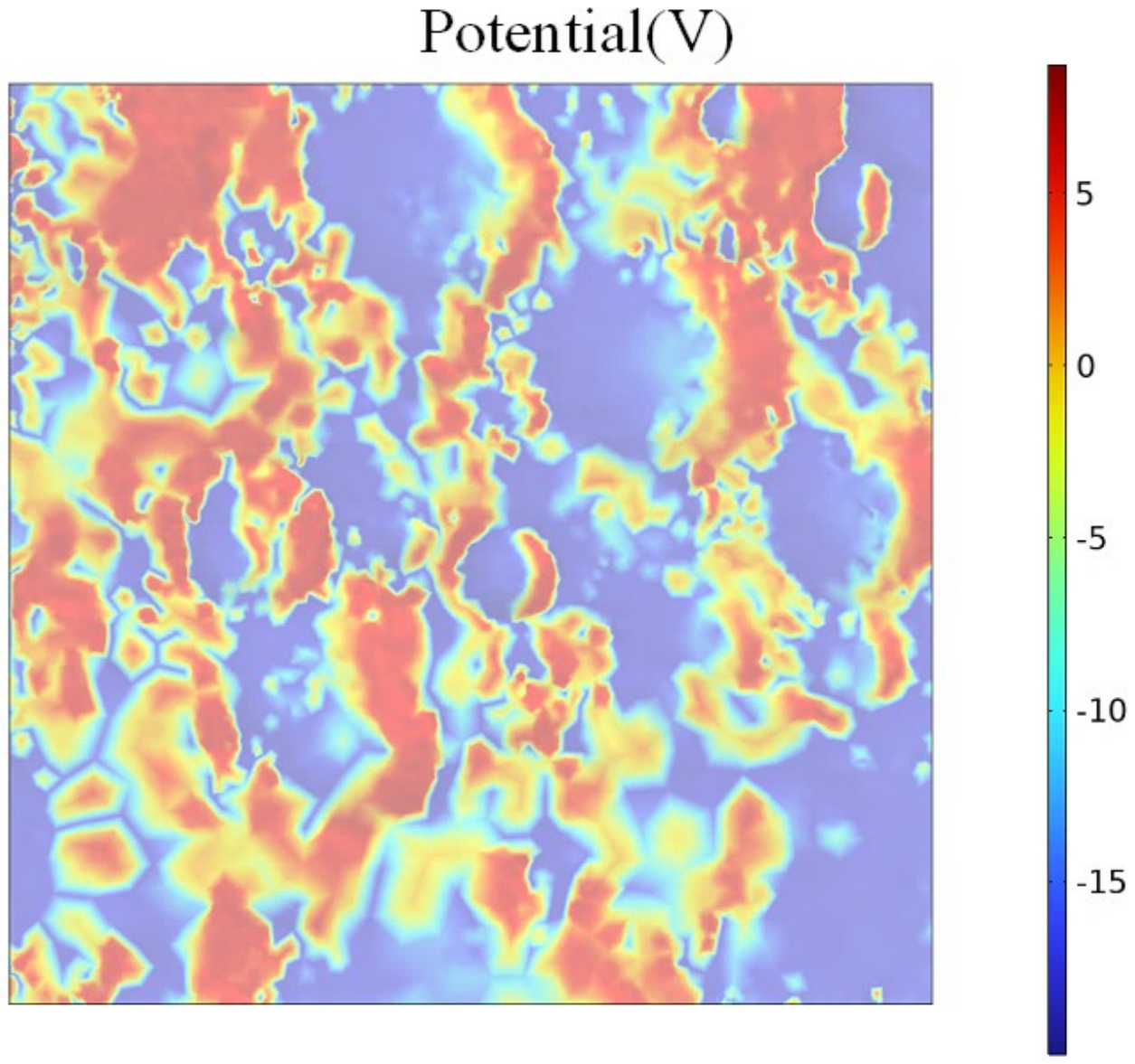}
        \caption{}
        \label{fig:8b}
    \end{subfigure}
    \begin{subfigure}{0.3\textwidth}
        \includegraphics[height=1.5in]{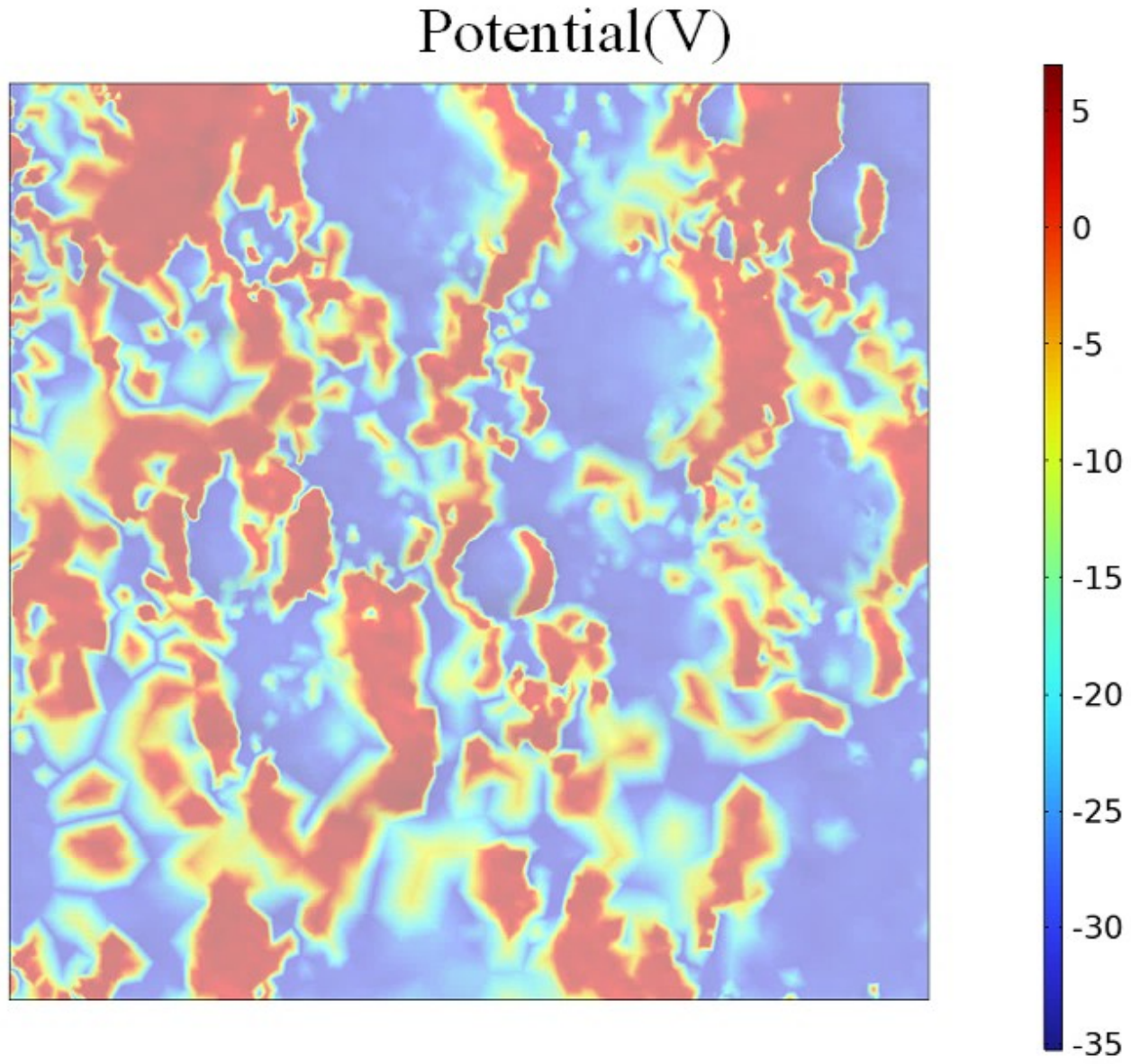}
        \caption{}
        \label{fig:8c}
    \end{subfigure}

    \vspace{0.01\textwidth}

    \begin{subfigure}{0.3\textwidth}
        \includegraphics[height=1.5in]{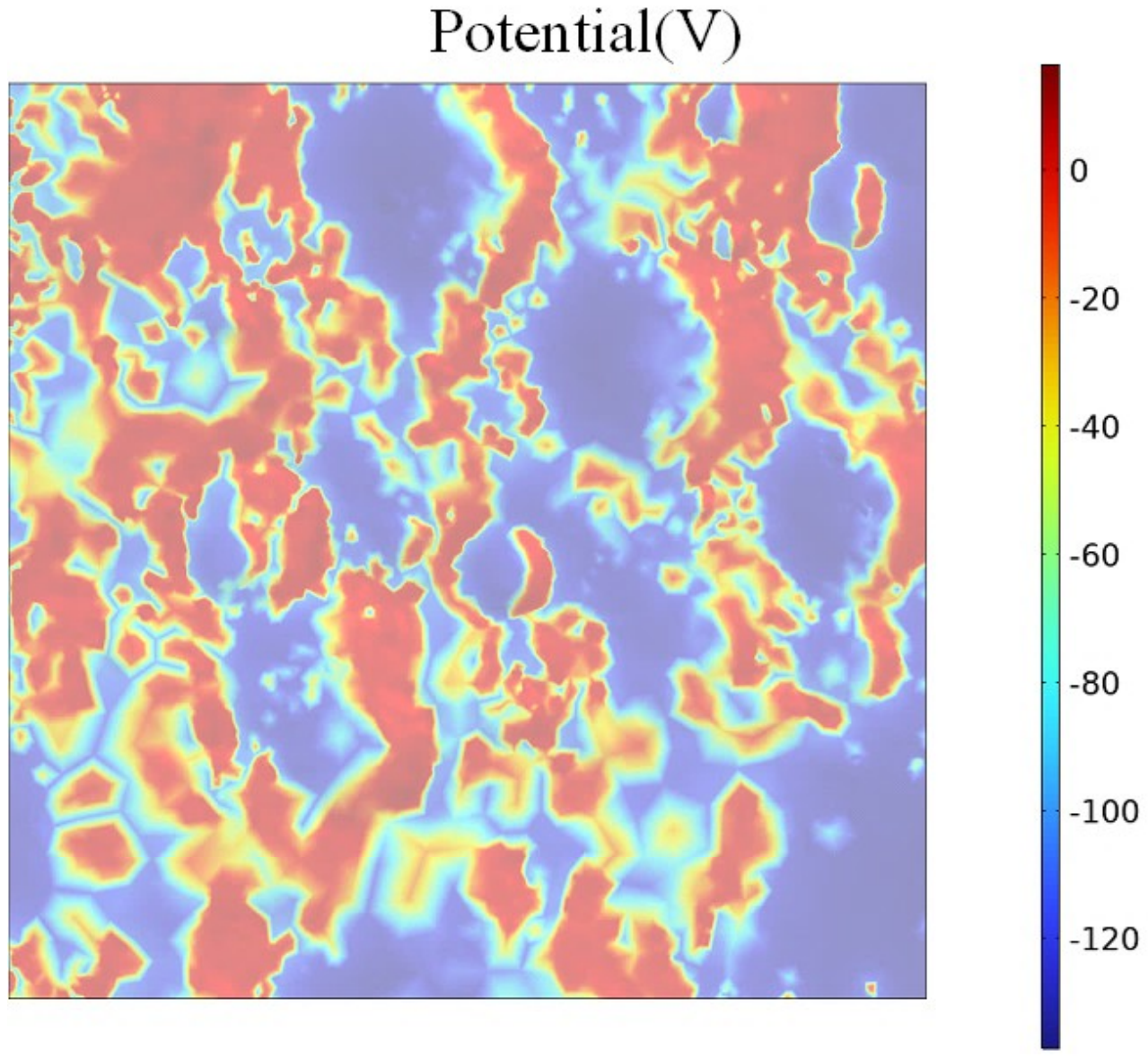}
        \caption{}
        \label{fig:8d}
    \end{subfigure}
    \begin{subfigure}{0.3\textwidth}
        \includegraphics[height=1.5in]{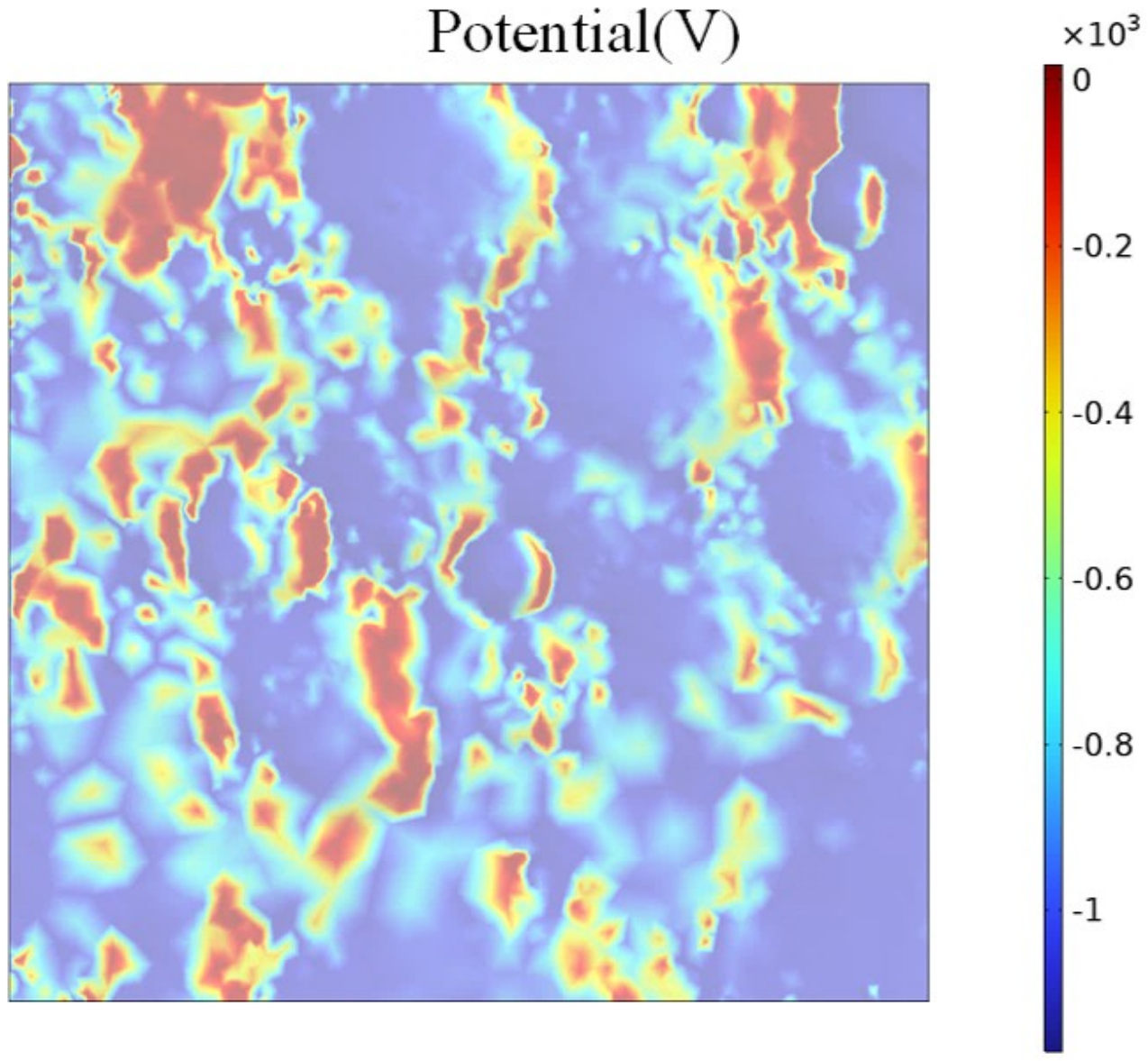}
        \caption{}
        \label{fig:8e}
    \end{subfigure}
    \begin{subfigure}{0.3\textwidth}
        \includegraphics[height=1.5in]{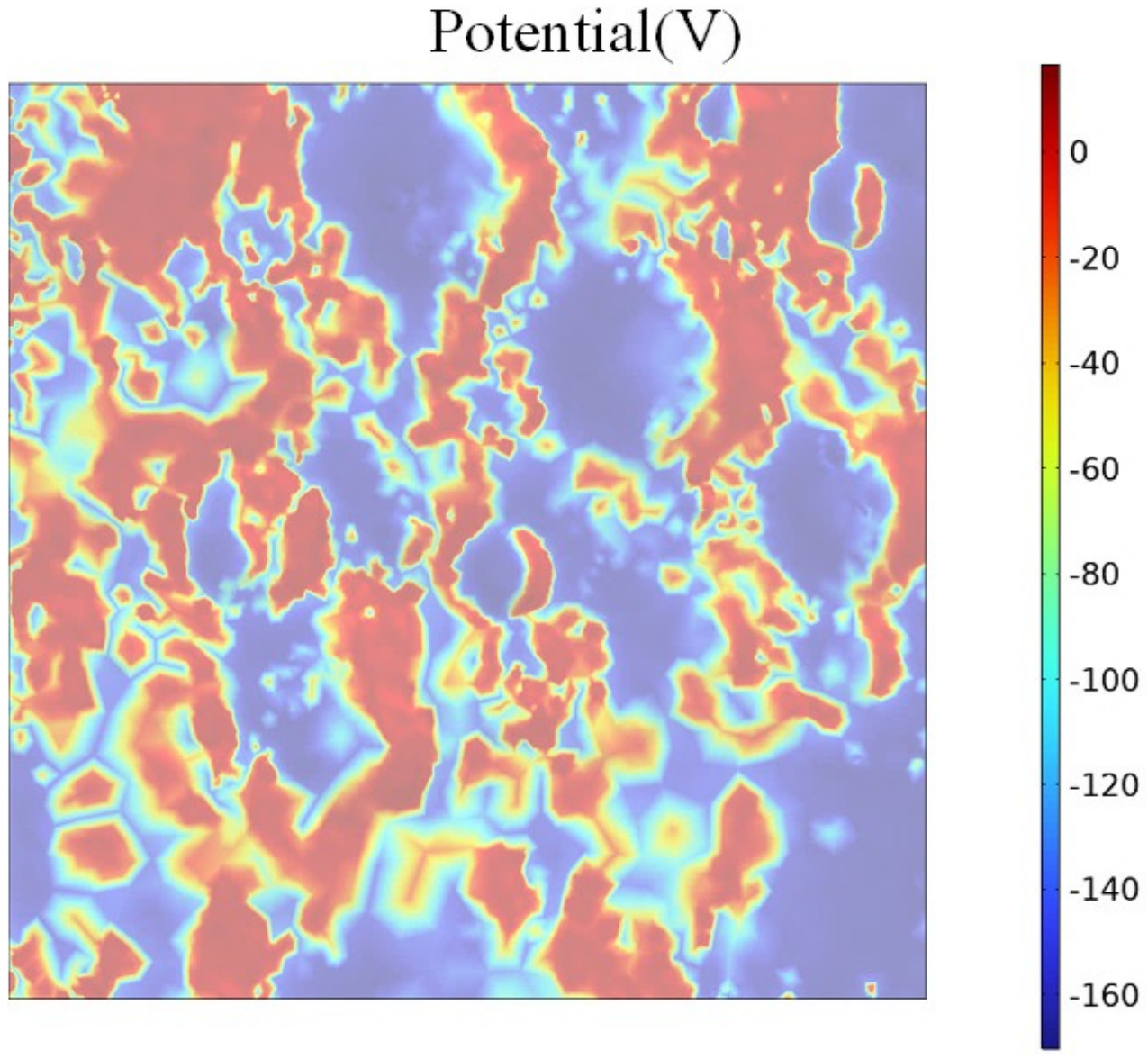}
        \caption{}
        \label{fig:8f}
    \end{subfigure}

    \vspace{0.01\textwidth}

    \begin{subfigure}{0.3\textwidth}
        \includegraphics[height=1.5in]{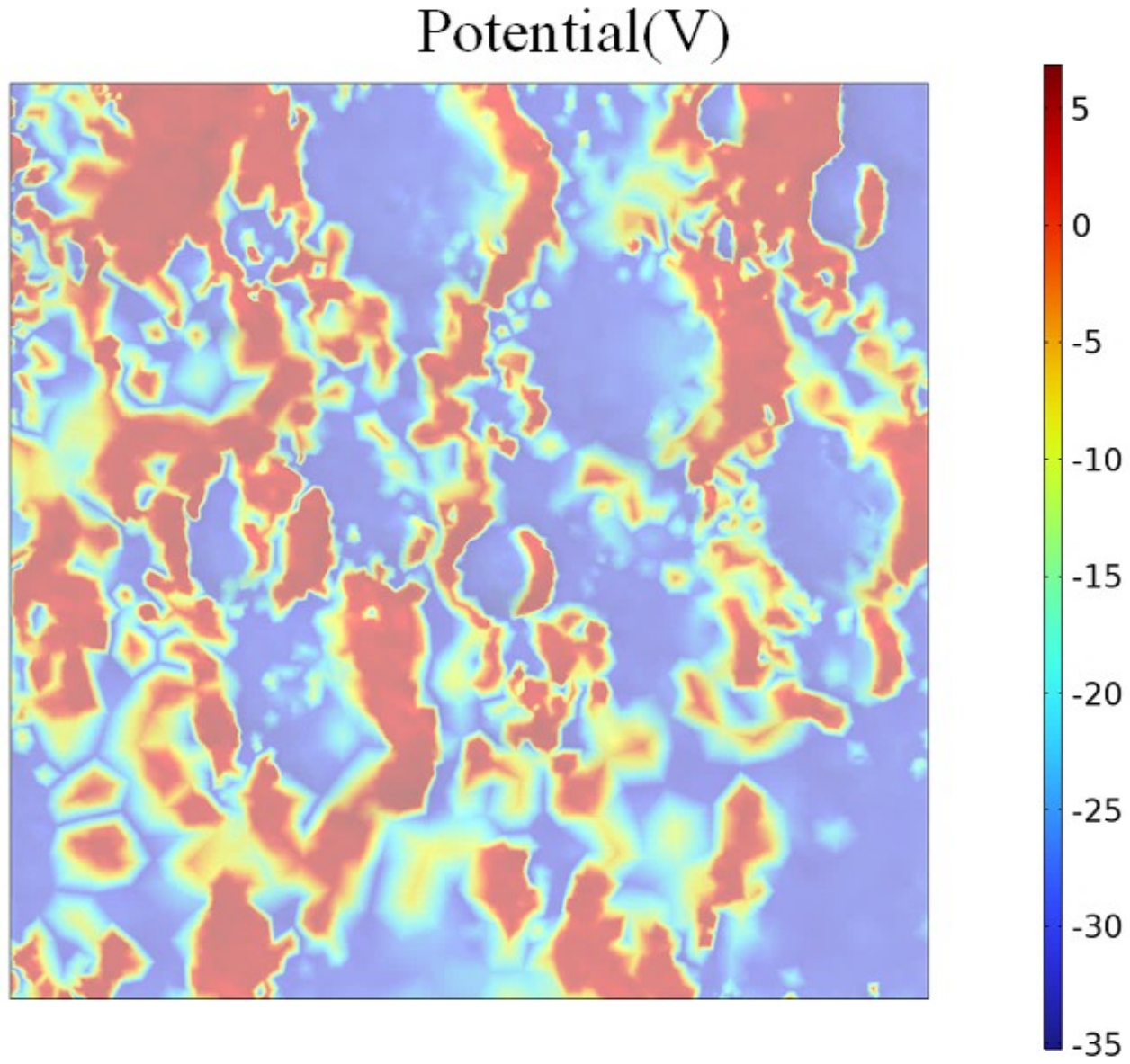}
        \caption{}
        \label{fig:8g}
    \end{subfigure}
    \begin{subfigure}{0.3\textwidth}
        \includegraphics[height=1.5in]{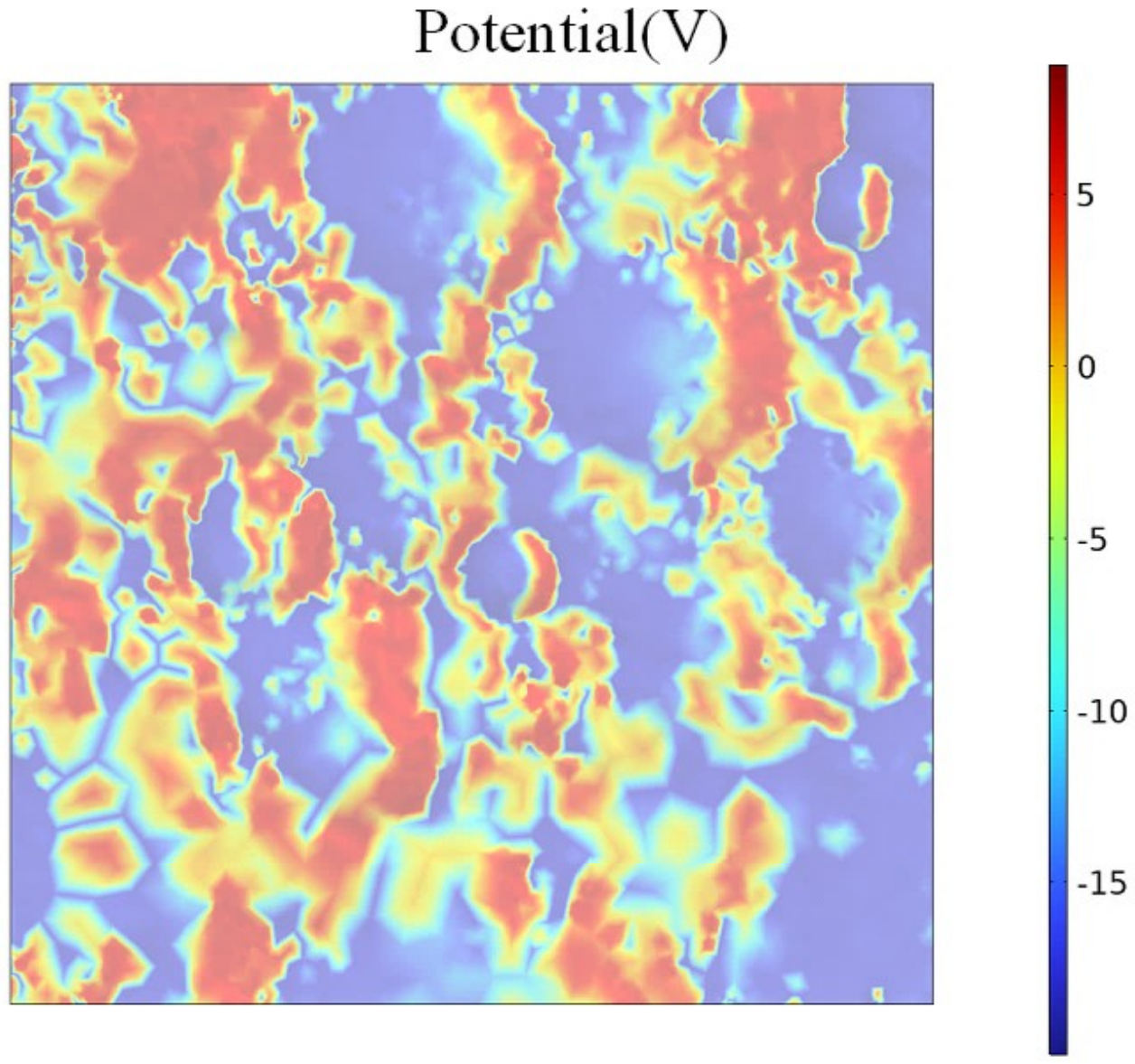}
        \caption{}
        \label{fig:8h}
    \end{subfigure}
    \begin{subfigure}{0.3\textwidth}
        \includegraphics[height=1.5in]{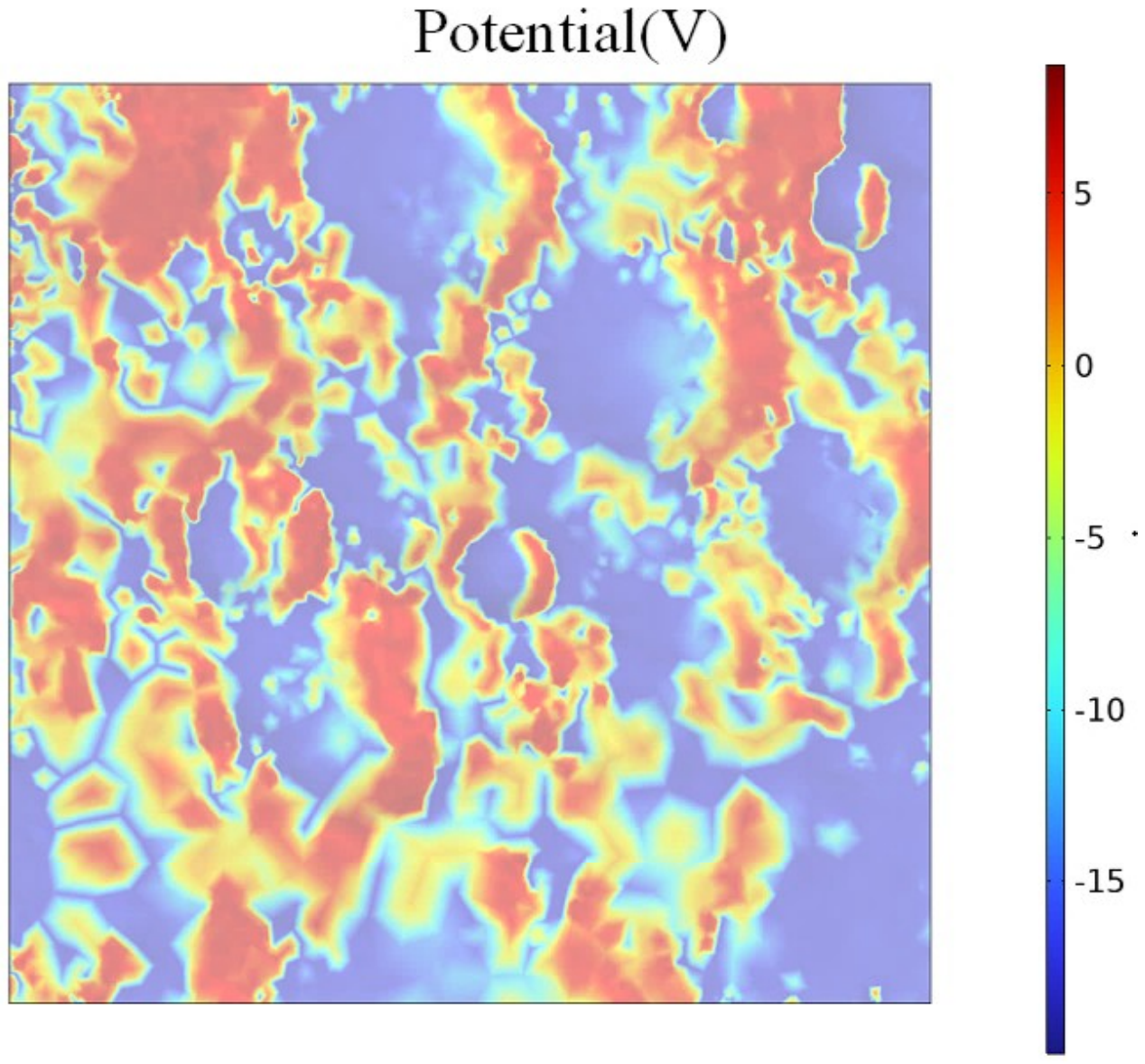}
        \caption{}
        \label{fig:8i}
    \end{subfigure}

    \caption{Surface potential (V) distributions at different lunar phases at a solar elevation angle of 1°.
    (\subref{fig:8a}) Surface potential distribution at a lunar phase of $-90^\circ$. (\subref{fig:8b}) Surface potential distribution at a lunar phase of $-67.5^\circ$. (\subref{fig:8c}) Surface potential distribution at a lunar phase of $-45^\circ$.
    (\subref{fig:8d}) Surface potential distribution at a lunar phase of $-22.5^\circ$.
    (\subref{fig:8e}) Surface potential distribution at a lunar phase of $0^\circ$.
    (\subref{fig:8f}) Surface potential distribution at a lunar phase of $22.5^\circ$.
    (\subref{fig:8g}) Surface potential distribution at a lunar phase of $45^\circ$.
    (\subref{fig:8h}) Surface potential distribution at a lunar phase of $67.5^\circ$.
    (\subref{fig:8i}) Surface potential distribution at a lunar phase of $90^\circ$.}
    \label{fig:8}
\end{figure}

With the solar elevation angle fixed at 1°, simulations were conducted to investigate the surface charging process of the lunar south pole under dynamically varying plasma conditions. To thoroughly analyze the surface charging characteristics of the lunar south pole in different regions of Earth’s magnetosphere, surface potential distributions of the lunar south pole are extracted at a lunar phase interval of 22.5° over half of the Moon’s orbital period, as presented in Figure 8(\subref{fig:8a})–(\subref{fig:8i}). The Moon is exposed to the solar wind during the intervals corresponding to Figure 8(\subref{fig:8a}), (\subref{fig:8b}), (\subref{fig:8h}), and (\subref{fig:8i}), is located in the magnetosheath during those corresponding to Figure 8(\subref{fig:8c}) and (\subref{fig:8g}), in the magnetotail lobes during those corresponding to Figure 8(\subref{fig:8d}) and (\subref{fig:8f}), and in the plasma sheet during that corresponding to Figure 8(\subref{fig:8e}).

As shown in the figures, in the solar wind environment, the minimum surface potential is approximately −20 V. In the magnetosheath, it is about −35 V. In the magnetotail lobe, it drops below −160 V. In the plasma sheet, the minimum potential even falls below −1000 V, accompanied by a notable expansion of the region with negative potential. 

According to the variations of plasma parameters, the plasma density is $7 \times 10^6\ \mathrm{m}^{-3}$ at the magnetosheath, which is higher than the value of $5 \times 10^6\ \mathrm{m}^{-3}$ under typical solar wind conditions. This leads to an increase in the flux of incident electrons and ions. As discussed previously, electron current contributes more significantly than ion current in topographic depressions. Consequently, the increased plasma density drives the minimum potential to more negative values. Meanwhile, the electron temperature and ion temperature rise to 26 eV and 80 eV respectively in the magnetosheath. Given that electrons have much smaller mass, their thermal velocity is more sensitive to temperature changes. The elevated electron temperature further shifts the surface potential toward negative values, whereas ion temperature exerts a relatively weak modulating effect.

At the magnetotail lobe, the plasma density decreases sharply to $3 \times 10^5\ \mathrm{m}^{-3}$, while the electron temperature and ion temperature increase to 180 eV and 540 eV. Although the reduced density lowers the flux of primary incident electrons, the remarkably high electron temperature greatly raises the thermal velocity of electrons and enhances the incident electron current. The influence of electron temperature dominates over the effect of density reduction, which further lowers the minimum surface potential. 

When the Moon enters the plasma sheet, the plasma density drops to an extremely low level of $1.33 \times 10^5\ \mathrm{m}^{-3}$, while the plasma temperature rises to as high as 2000 eV. The extremely high temperature endows electrons with strong thermal motion. Even under low plasma density, a large number of high-energy electrons can still overcome the potential barrier of the negatively charged lunar surface and continuously reach the surface. To achieve dynamic balance of particle currents, the lunar surface potential has to decrease to an extremely low level to block incoming electrons, which is the primary reason why the surface potential falls below −1000 V in the plasma sheet.

\begin{figure}[!ht]
    \centering
    \begin{subfigure}{0.3\textwidth}
        \includegraphics[height=1.5in]{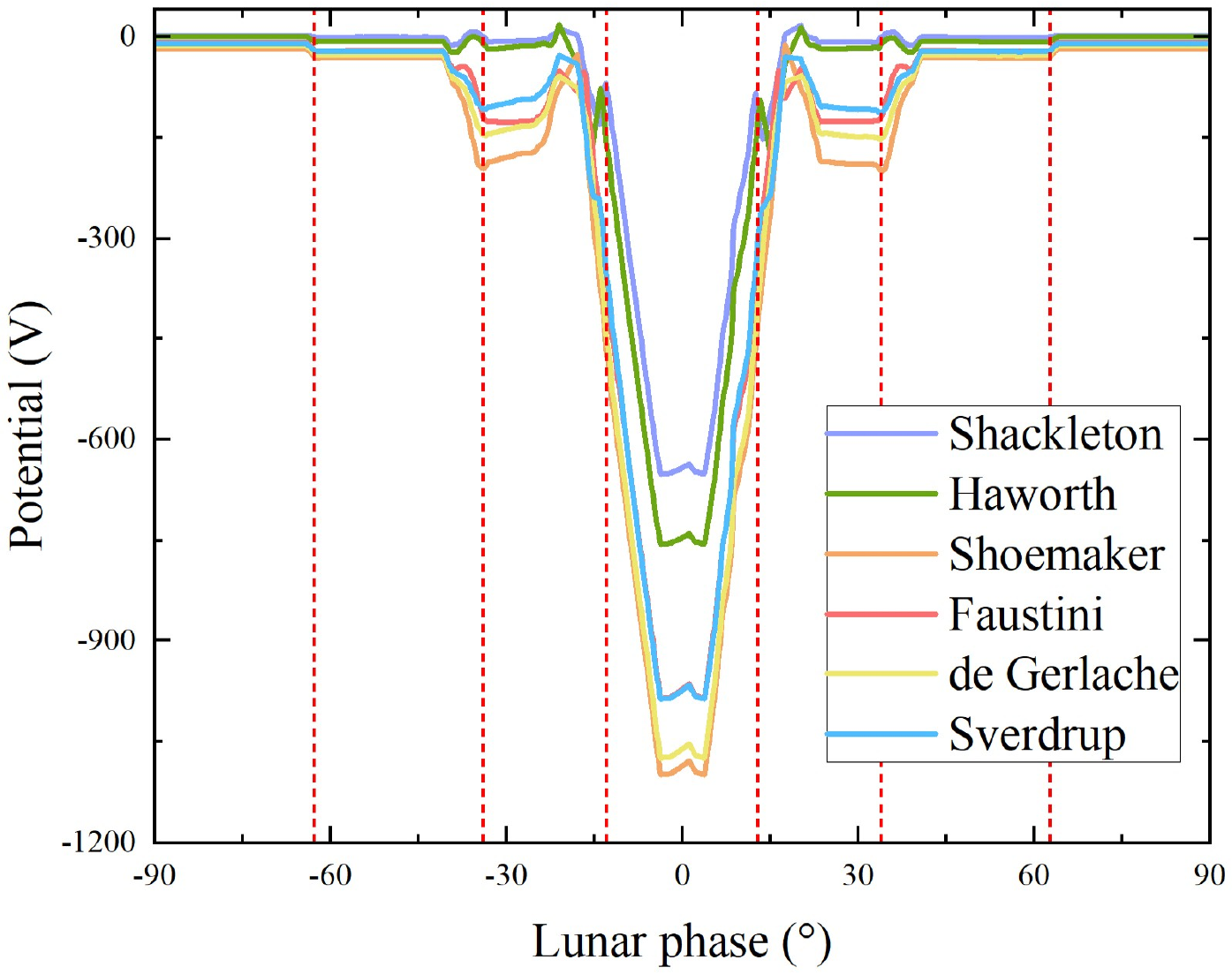}
        \caption{}
        \label{fig:9a}
    \end{subfigure}
    \begin{subfigure}{0.3\textwidth}
        \includegraphics[height=1.5in]{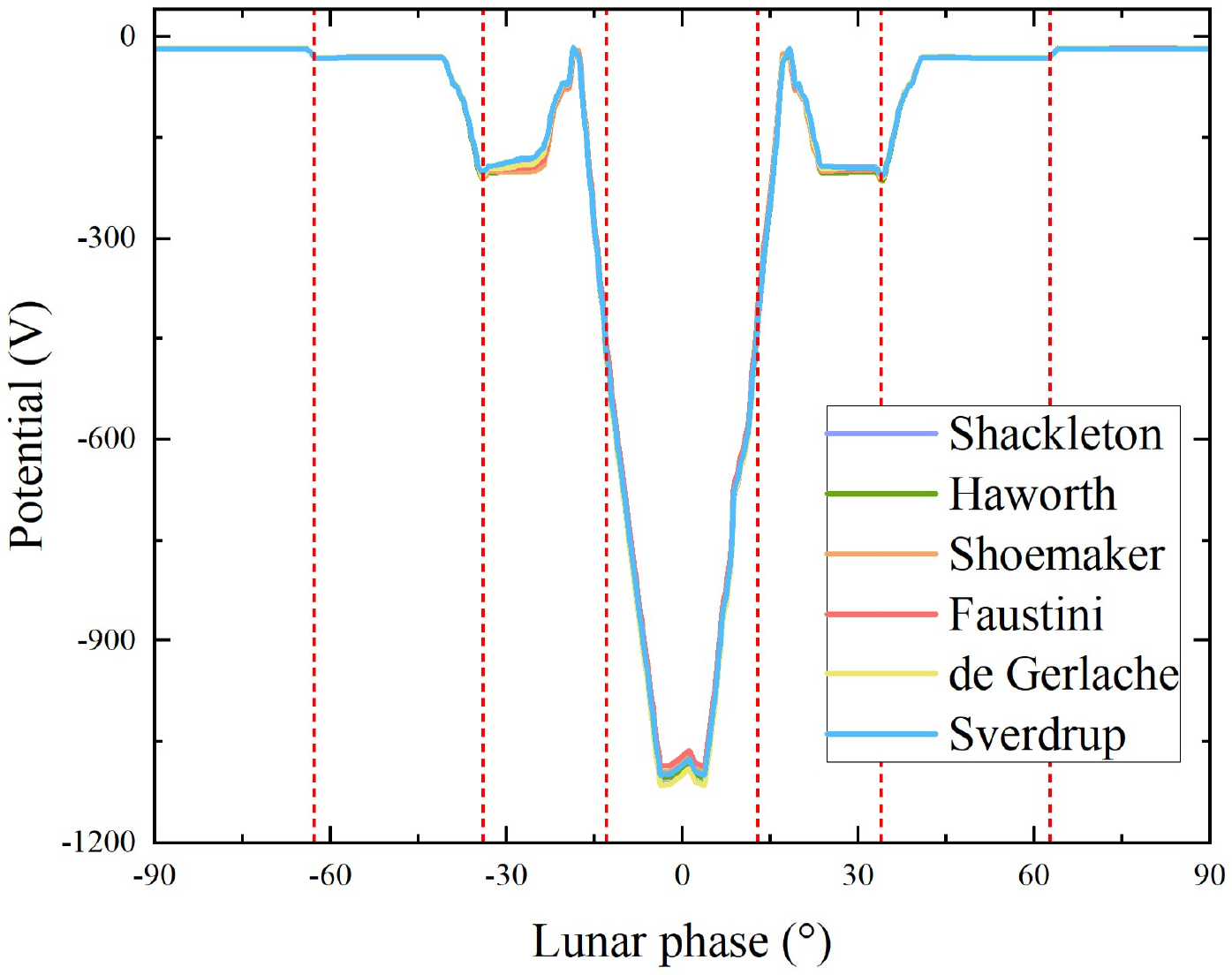}
        \caption{}
        \label{fig:9b}
    \end{subfigure}
    \begin{subfigure}{0.3\textwidth}
        \includegraphics[height=1.5in]{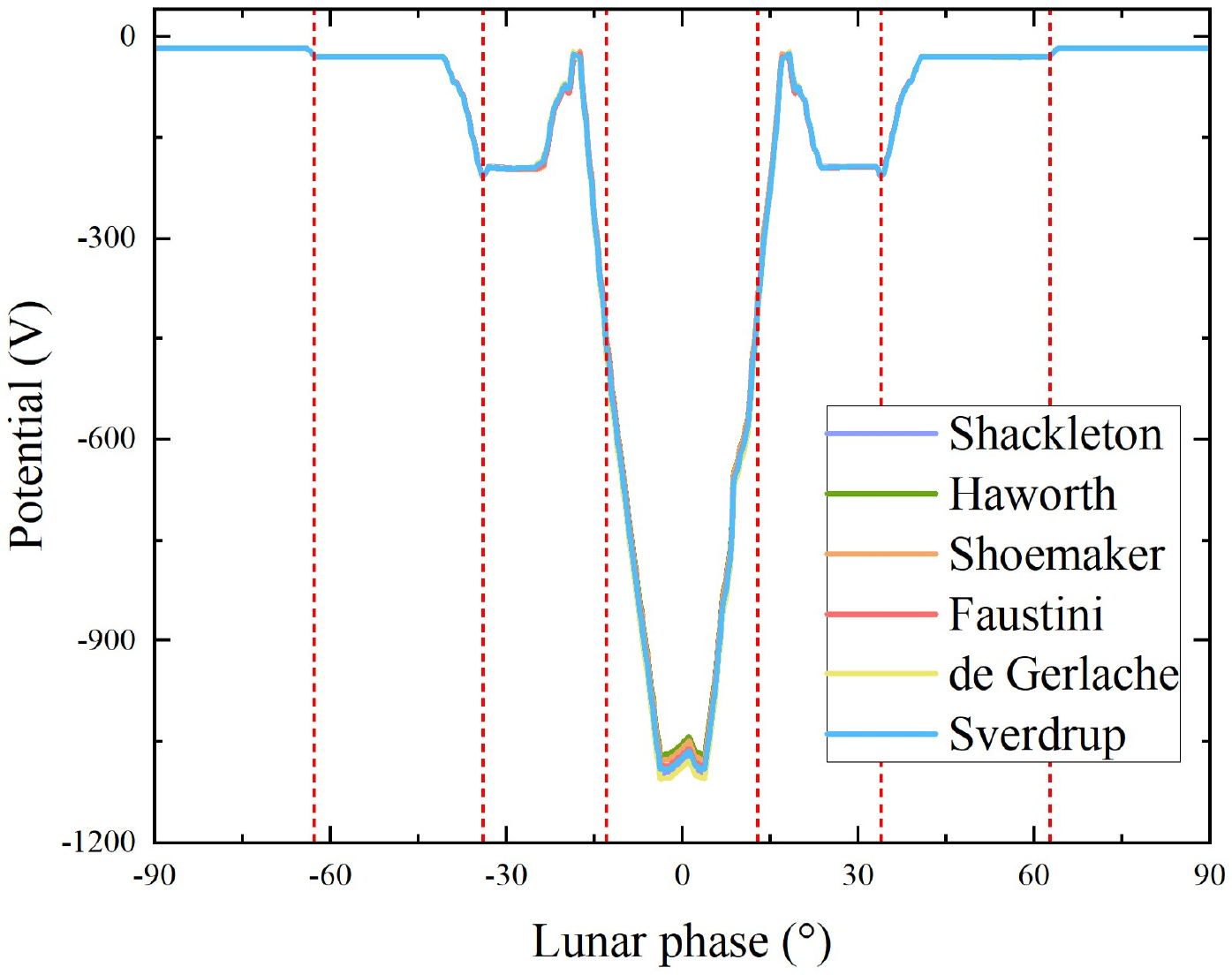}
        \caption{}
        \label{fig:9c}
    \end{subfigure}

    \vspace{0.03\textwidth}

    \begin{subfigure}{0.3\textwidth}
        \includegraphics[height=1.5in]{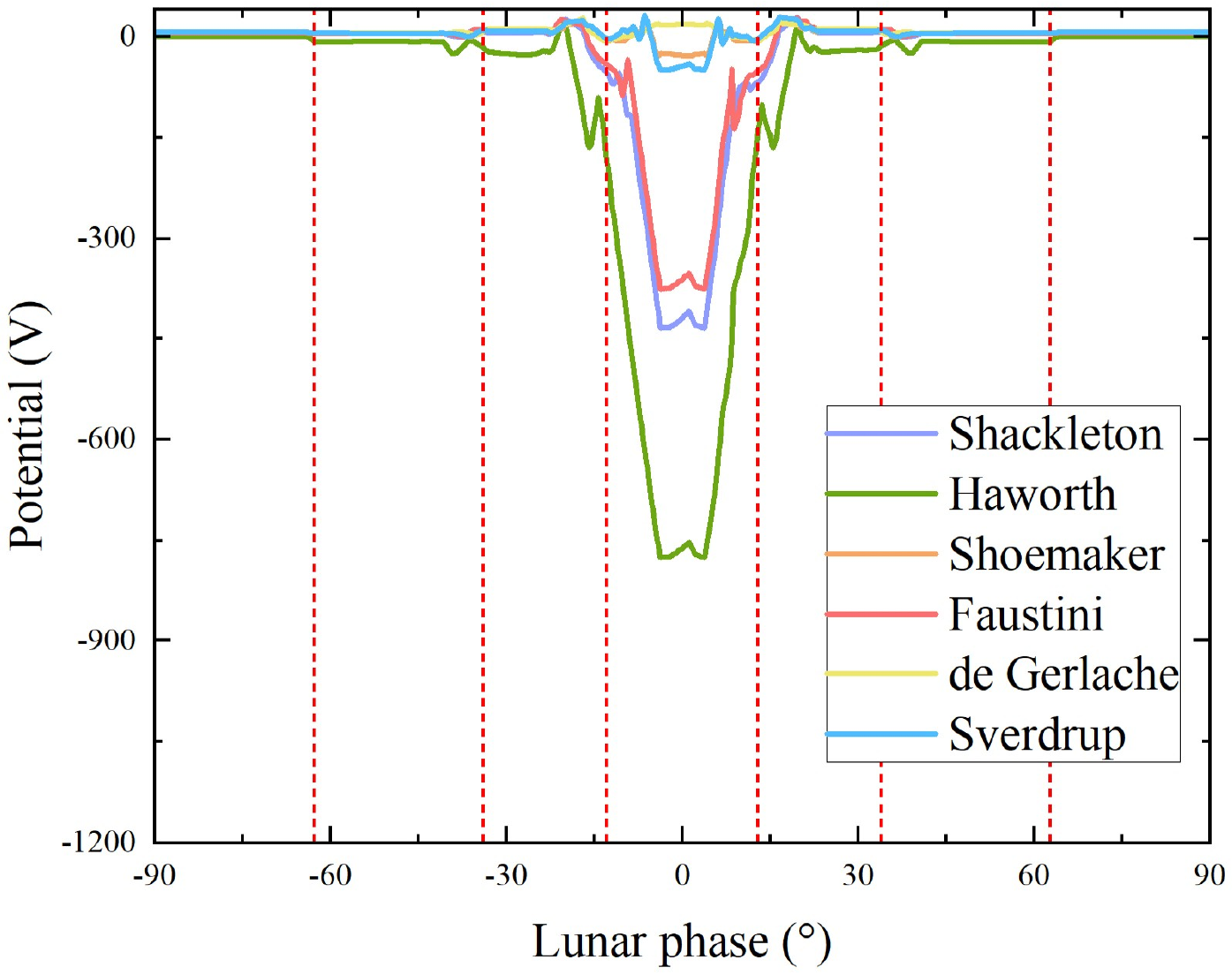}
        \caption{}
        \label{fig:9d}
    \end{subfigure}
    \begin{subfigure}{0.3\textwidth}
        \includegraphics[height=1.5in]{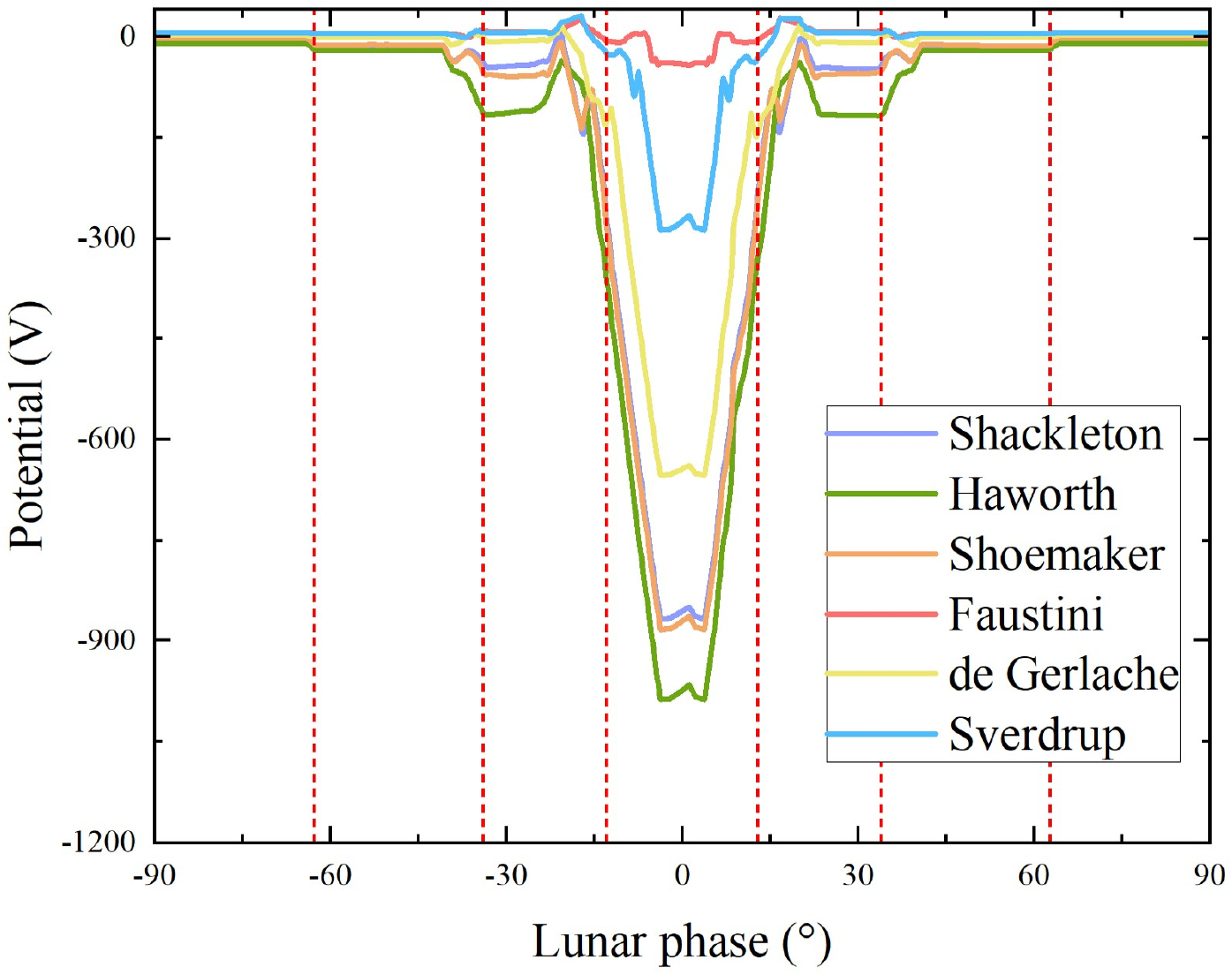}
        \caption{}
        \label{fig:9e}
    \end{subfigure}
    \begin{subfigure}{0.3\textwidth}
        \includegraphics[height=1.5in]{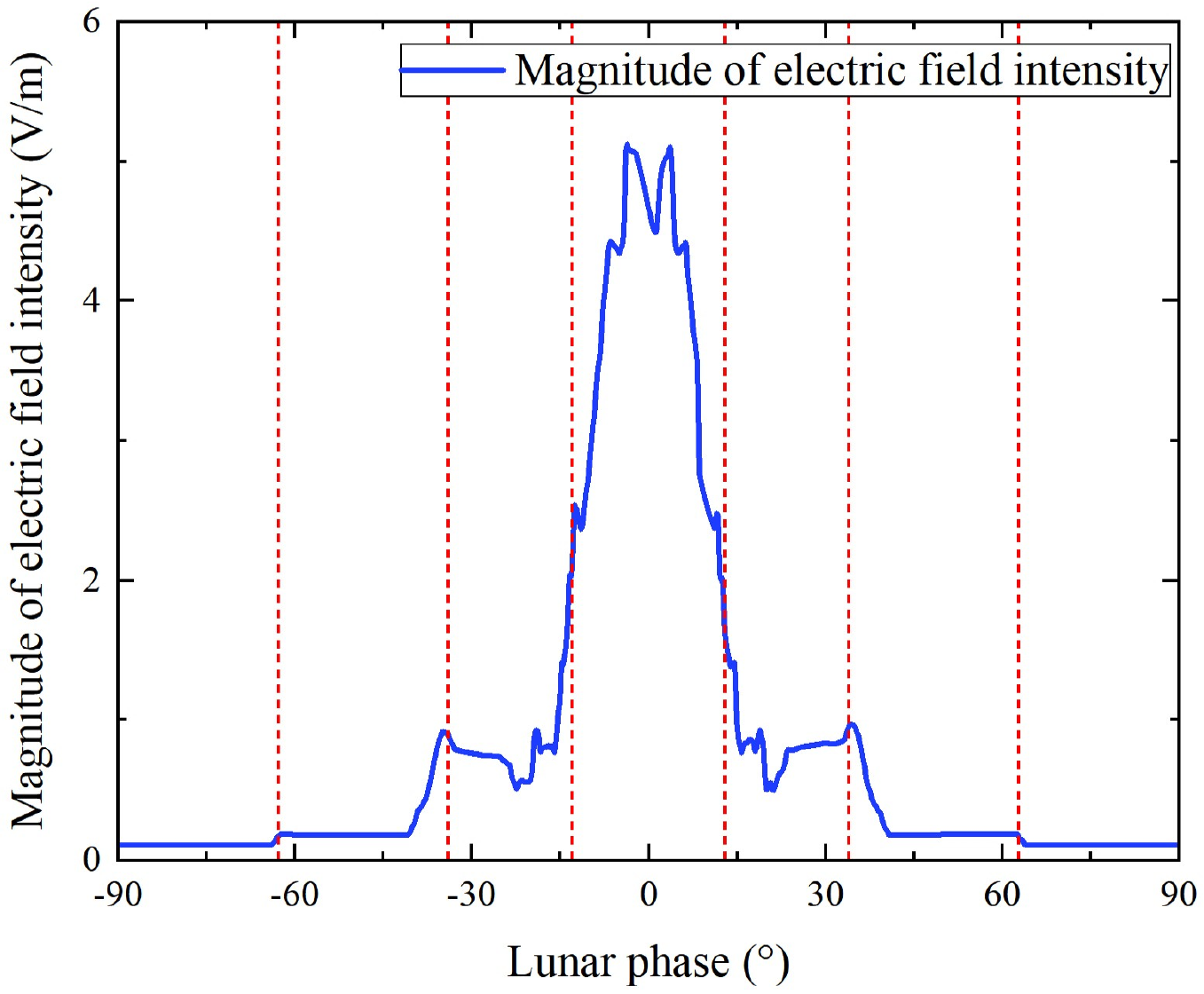}
        \caption{}
        \label{fig:9f}
    \end{subfigure}

   \caption{(\subref{fig:9a}) Curves of potential (V) of the top of the upstream crater wall versus lunar phase for six typical craters. (\subref{fig:9b}) Curves of potential (V) of the middle of the upstream crater wall versus lunar phase for six typical craters. (\subref{fig:9c}) Curves of potential (V) of the crater floor versus lunar phase for six typical craters. (\subref{fig:9d}) Curves of potential (V) of the middle of the downstream crater wall versus lunar phase for six typical craters. (\subref{fig:9e}) Curves of potential (V) of the top of the downstream crater wall versus lunar phase for six typical craters.(\subref{fig:9f}) Curve of maximum electric field magnitude (V/m) versus lunar phase.}
    \label{fig:9}
\end{figure}

Figure 9(\subref{fig:9a})–(\subref{fig:9e}) present the potential variations with lunar phase over half of the lunar orbital period at five representative locations in six typical craters near the lunar south pole: the top of the upstream crater wall, the middle of the upstream crater wall, the crater floor, the middle of the downstream crater wall, and the top of the downstream crater wall.

Overall, the potential curves at all locations are approximately symmetric about the lunar phase of 0°. Among the five locations, the curves for the middle of the upstream crater wall and the crater floor nearly overlap among the six craters, and their potentials are generally lower than those at the other locations. This is mainly because these two regions remain shadowed for long periods and lie within the plasma cavity. By contrast, the middle of the downstream crater wall is more readily illuminated by sunlight, leading to relatively higher potentials.

In addition, pronounced differences are observed among the potential curves at the crater-wall tops and at the middle of the downstream crater wall for different craters, indicating that the surface charging process in these regions is more sensitive to local topographic conditions. These differences are mainly caused by the fact that the crater walls are not ideal smooth surfaces, but instead exhibit varying degrees of roughness and slope variation. Local protrusions or depressions can modify the solar illumination conditions and the incident plasma flux, thereby affecting photoelectron emission and plasma collection processes. This effect is particularly important near the crater-wall tops, which are usually located in the transition region between illuminated and shadowed areas, where the potential varies rapidly. As a result, even small topographic differences may lead to noticeable potential deviations. In some cases, parts of the upstream crater wall may also be shielded by nearby elevated terrain and located in a local wake region, causing them to exhibit charging characteristics different from those of other craters at the same lunar phase.

Although the potential values differ among the selected locations and craters, most of the potential curves exhibit an overall downward trend from a lunar phase of −90° to 0°, followed by an overall upward trend thereafter. Focusing on the first half of the potential curves at the crater floor, the crater-floor potential remains stable at approximately −17 V under typical solar wind conditions. It decreases to around −30 V in the magnetosheath and drops rapidly to about −200 V in the front part of the magnetotail lobe. In the rear part of the magnetotail lobe, the potential stops decreasing and rises noticeably to a level similar to that in the magnetosheath, forming distinct fluctuations. The potential begins to fall again when approaching the plasma sheet, keeps decreasing in the early part of the plasma sheet, and finally levels off at around −1000 V in the late stage of this region.

The potential fluctuations in the magnetotail lobe can be explained by plasma parameter changes. The plasma density is relatively high in the front part of the magnetotail lobe but decreases to a very low level in the rear part. Meanwhile, the rising electron temperature enhances the secondary electron emission from lunar regolith. Since the electron temperature here is not as extreme as that in the plasma sheet, the effect of secondary electron emission outweighs the increased electron current caused by temperature rise, thereby resulting in the recovery of surface potential.

Figure 9(\subref{fig:9f}) displays the evolution of the maximum magnitude of the electric field across the entire simulation domain. The curve is also approximately symmetric about the lunar phase of 0°. Focusing on the first half of the curve, the magnitude of the electric field shows an overall upward trend. It is about 0.1 V/m under typical solar wind conditions, rises to around 0.18 V/m in the magnetosheath and approximately 0.75 V/m in the magnetotail lobe, and finally reaches nearly 5 V/m in the plasma sheet. The lunar surface consists of sunlit regions and permanently shadowed regions. Photoelectron emission maintains weak positive potential or near-zero potential in sunlit areas, while continuous negative charging occurs in permanently shadowed regions. As the absolute value of negative potential in shadowed areas increases, the potential gradient between sunlit and shadowed regions grows gradually, which accounts for the rising trend of the electric field. In addition, the slight recovery of potential in the rear magnetotail lobe also leads to temporary fluctuations in the electric field.

\section{Conclusion}

This study adopts the digital elevation model (DEM) retrieved from observations by the Lunar Orbiter Laser Altimeter (LOLA) to establish a geometric terrain model of the lunar south pole. Combining the finite element method with the BP neural network, an efficient multi-physics simulation model for surface charging at the lunar south pole is constructed. Using this model, we simulate the surface charging process under different solar altitude angles, and carry out long-term dynamic simulations of charging characteristics as the Moon travels sequentially through the typical solar wind and different plasma regions of Earth’s magnetosphere.

The results show that the surface charging process in the lunar south pole is jointly controlled by topography, solar elevation angle, and dynamic plasma environment parameters. Under extremely low solar elevation angle conditions, the lunar surface potential distribution exhibits a clear dependence on local topography: windward slopes of raised landforms tend to reach relatively higher potentials, whereas crater floors, leeward slopes, and strongly shielded regions are more negatively charged due to reduced photoelectron emission and obstructed plasma flow. As a result, relatively sharp potential variations occur near raised landforms and at the junctions between crater floors and downstream crater walls, producing enhanced local electric fields. In addition, pronounced differences are observed among different craters in the potential curves at the crater-wall tops and the middle of the downstream crater wall, suggesting that surface charging in these regions is more sensitive to local topographic conditions.

Under typical solar wind conditions, when the solar elevation angle increases from 0° to 2°, the topographic shielding effect weakens. The area with positive surface potential expands gradually, and the boundary between positive and negative potential inside craters shifts toward the upstream crater walls. The maximum surface potential increases from 8.24 V to 8.75 V, and the minimum surface potential increases from −20.05 V to −19.47 V, suggesting that slight variations in the solar elevation angle exert a stronger influence on sunlit highland regions.

During the dynamic variation of plasma parameters, the potential and electric-field curves are approximately symmetric about the lunar phase of 0°. As the Moon successively passes through the solar wind, magnetosheath, magnetotail lobe, and plasma sheet, the surface potential generally decreases. An exception occurs in the latter half of the magnetotail-lobe interval, where the surface potential shows a clear rebound. This is because the plasma density decreases significantly, while the electron temperature has not yet reached the extremely high level characteristic of the plasma sheet. As a result, the enhanced secondary electron emission exceeds the increase in incident electron current caused by the rising electron temperature, leading to the recovery of the surface potential. Under typical solar wind conditions, the crater-floor potential remains nearly stable at approximately −17 V. After entering the magnetosheath, it decreases to about −30 V, and then rapidly drops to approximately −200 V in the early stage of the magnetotail-lobe region. Upon entering the plasma sheet, the potential decreases sharply again and stabilizes at approximately −1000 V, accompanied by a marked expansion of the negative-potential region. The maximum magnitude of the electric field over the entire simulation domain also increases from about 0.1 V/m in the solar wind environment to about 5 V/m in the plasma sheet. The extremely negative potential and strong electric field in the plasma sheet may raise the risk of electrostatic discharge on lunar surface equipment, affect the charging and migration of lunar dust, and threaten the operational safety of lunar detectors and long-term surface facilities.

The research findings can provide references for landing site selection, rover path planning, anti-static design of lunar surface equipment, and space environment safety assessment for future lunar bases.

\section*{Acknowledgments}

\subsection*{Author Contributions} 
Y. Xiao: Conceptualization, Methodology, Software, Data curation, Writing – original draft.
X. Wang: Investigation, Data curation, Writing – review \& editing.
R. Quan: Methodology, Software, Writing – review \& editing.

\subsection*{Funding}
This work was supported by the National Natural Science Foundation of China [grant number 42241148]. 

\subsection*{Conflicts of Interest}
The authors declare that there is no conflict of interest regarding the publication of this article.

\subsection*{Data Availability}
All data supporting the findings of this study will be made available from the corresponding author on reasonable request.

\printbibliography

@article{Halekas1,
   author = {Halekas, J. S. and Delory, G. T. and Brain, D. A.},
   title = {Extreme lunar surface charging during solar energetic particle events},
   journal = {Geophysical Research Letters},
   volume = {34},
   number = {2},
   year = {2007}
}

@article{Poppe1,
   author = {Poppe, A. R. and Halekas, J. S. and Delory, G. T.},
   title = {A comparison of ARTEMIS observations and particle-in-cell modeling of the lunar photoelectron sheath in the terrestrial magnetotail},
   journal = {Geophysical Research Letters},
   volume = {39},
   number = {1},
   year = {2012}
}

@article{Harada1,
   author = {Harada, Y. and Poppe, A. R. and Halekas, J. S.},
   title = {Photoemission and electrostatic potentials on the dayside lunar surface in the terrestrial magnetotail lobes},
   journal = {Geophysical Research Letters},
   volume = {44},
   number = {11},
   pages = {5276--5282},
   year = {2017}
}

@article{Nishino1,
   author = {Nishino, M. N. and Harada, Y. and Saito, Y.},
   title = {Kaguya observations of the lunar wake in the terrestrial foreshock: Surface potential change by bow-shock reflected ions},
   journal = {Icarus},
   volume = {293},
   pages = {45--51},
   year = {2017}
}

@article{Farrell2,
   author = {Farrell, W. M. and Stubbs, T. J. and Vondrak, R. R.},
   title = {Complex electric fields near the lunar terminator: The near-surface wake and accelerated dust},
   journal = {Geophysical Research Letters},
   volume = {34},
   number = {14},
   year = {2007}
}

@article{Halekas2,
   author = {Halekas, J. S. and Delory, G. T. and Lin, R. P.},
   title = {Lunar Prospector measurements of secondary electron emission from lunar regolith},
   journal = {Planetary and Space Science},
   volume = {57},
   number = {1},
   pages = {78--82},
   year = {2009}
}

@article{Stubbs1,
   author = {Stubbs, T. J. and Farrell, W. M. and Halekas, J. S.},
   title = {Dependence of lunar surface charging on solar wind plasma conditions and solar irradiation},
   journal = {Planetary and Space Science},
   volume = {90},
   pages = {10--27},
   year = {2014}
}

@article{Halekas3,
   author = {Halekas, J. S. and Delory, G. T. and Lin, R. P.},
   title = {Lunar Prospector observations of the electrostatic potential of the lunar surface and its response to incident currents},
   journal = {Journal of Geophysical Research: Space Physics},
   volume = {113},
   number = {A9},
   year = {2008}
}

@article{Halekas4,
   author = {Halekas, J. S. and Delory, G. T. and Farrell, W. M.},
   title = {First remote measurements of lunar surface charging from ARTEMIS: Evidence for nonmonotonic sheath potentials above the dayside surface},
   journal = {Journal of Geophysical Research: Space Physics},
   volume = {116},
   number = {A7},
   year = {2011}
}

@article{Harada2,
   author = {Harada, Y. and Machida, S. and Halekas, J. S.},
   title = {ARTEMIS observations of lunar dayside plasma in the terrestrial magnetotail lobe},
   journal = {Journal of Geophysical Research: Space Physics},
   volume = {118},
   number = {6},
   pages = {3042--3054},
   year = {2013}
}

@article{Poppe2,
   author = {Poppe, A. R. and Xu, S. and Liuzzo, L.},
   title = {ARTEMIS observations of lunar nightside surface potentials in the magnetotail lobes: Evidence for micrometeoroid impact charging},
   journal = {Geophysical Research Letters},
   volume = {48},
   number = {15},
   year = {2021}
}

@article{LiS1,
   author = {Li, S. and Poppe, A. R. and Orlando, T. M.},
   title = {Formation of lunar surface water associated with high-energy electrons in Earth’s magnetotail},
   journal = {Nature Astronomy},
   volume = {7},
   pages = {1427--1435},
   year = {2023}
}

@article{Wang1,
   author = {Wang, W. and Jin, Q. and Chen, X.},
   title = {Character and spatial distribution of mineralogy at the lunar south polar region},
   journal = {Planetary and Space Science},
   volume = {240},
   pages = {105833},
   year = {2024}
}

@article{Farrell3,
   author = {Farrell, W. M. and Stubbs, T. J. and Halekas, J. S.},
   title = {Loss of solar wind plasma neutrality and affect on surface potentials near the lunar terminator and shadowed polar regions},
   journal = {Geophysical Research Letters},
   volume = {35},
   number = {5},
   year = {2008}
}

@article{Farrell1,
   author = {Farrell, W. M. and Stubbs, T. J. and Halekas, J. S.},
   title = {Anticipated electrical environment within permanently shadowed lunar craters},
   journal = {Journal of Geophysical Research: Planets},
   volume = {115},
   year = {2010}
}

@article{Zimmerman1,
   author = {Zimmerman, M. I. and Jackson, T. L. and Farrell, W. M.},
   title = {Plasma wake simulations and object charging in a shadowed lunar crater during a solar storm},
   journal = {Journal of Geophysical Research: Planets},
   volume = {117},
   year = {2012}
}

@inproceedings{Lund1,
  author = {Lund, D. and Zhao, J. and Lamb, A.},
  title = {Fully kinetic pife-pic simulations of plasma charging at lunar craters},
  booktitle = {AIAA Scitech 2020 Forum},
  year = {2020},
  pages = {1549}
}

@article{Xia1,
   author = {Xia, Q. and Cai, M. H. and Xu, L. L.},
   title = {Distribution of charged lunar dust in the south polar region of the moon},
   journal = {Chinese Physics B},
   volume = {31},
   pages = {045201},
   year = {2022}
}

@article{Gan1,
  author  = {Gan, H. and Zhao, C. and Wei, G.},
  title   = {Numerical simulation of the lunar polar environment: Implications for rover exploration challenge},
  journal = {Aerospace},
  volume  = {10},
  pages   = {598},
  year    = {2023}
}

@article{Zhao1,
  author  = {Zhao, C. and Gan, H. and Xie, L.},
  title   = {Theoretical analysis of the electric potential and the electrostatic dust transport around the Shackleton crater in the Lunar south pole region},
  journal = {Science China Earth Sciences},
  volume  = {66},
  pages   = {2278--2286},
  year    = {2023}
}

@article{Barker1,
  author  = {Barker, M. K. and Mazarico, E. and Neumann, G. A.},
  title   = {A new view of the lunar south pole from the lunar orbiter laser altimeter (LOLA)},
  journal = {The Planetary Science Journal},
  volume  = {4},
  pages   = {183},
  year    = {2023}
}

@misc{Stopar1,
  author       = {Stopar, J. and Meyer, H.},
  title        = {Topographic Map of the Moon's South Pole (85°S to Pole)},
  organization = {Lunar and Planetary Institute Regional Planetary Image Facility},
  note         = {LPI Contribution 2171},
  year         = {2019}
}

@article{ZhangH1,
   author = {Zhang, H. and Quan, R. and Zhang, C.},
   title = {Inversion Analysis of GEO Plasma Environmental Parameters Based on BP Neural Network},
   journal = {Chinese Journal of Space Science},
   volume = {43},
   number = {1},
   pages = {78--86},
   year = {2023}
}

@article{LiuH1,
   author = {Liu, H. and Xu, Y. and Wang, C.},
   title = {Study on the linkages between microstructure and permeability of porous media using pore network and BP neural network},
   journal = {Materials Research Express},
   volume = {9},
   number = {2},
   year = {2022}
}

@article{song1,
  title={Three-dimensional simulation of surface charging in meteorite craters on rotating asteroids},
  author={Song, Z. Y. and Liu, Z. G. and Quan, R. H.},
  journal={Planetary and Space Science},
  volume={260},
  pages={106089},
  year={2025}
}

@article{pandya1,
  author    = {Pandya, A. and Mehta, P. and Kothari, N.},
  title     = {Impact of secondary and backscattered electron currents on absolute charging of structures used in spacecraft},
  journal   = {International Journal of Numerical Modelling: Electronic Networks, Devices and Fields},
  year      = {2019},
  volume    = {32},
  pages     = {e2631}
}

@article{wang2,
  author    = {Wang, S. and Wu, Z. C. and Tang, X. J. and Yi, Z.},
  title     = {A new charging model for spacecraft exposed dielectric (SICCE)},
  journal   = {IEEE Transactions on Plasma Science},
  year      = {2016},
  volume    = {44},
  pages     = {289--295}
}

@article{derosa1,
  title={Characterisation of potential landing sites for the European Space Agency's Lunar Lander project},
  author={De Rosa, D. and Bussey, B. and Cahill, J.T.},
  journal={Planetary and space science},
  volume={74},
  pages={224--246},
  year={2012},
}

@article{zeng1,
  title={Study on Lunar Surface Charging Effects Induced by Charged Particle Flows in the Earth’s Magnetotail Lobes and Spatial Distribution Characteristics of Charged Lunar Dust},
  author={Zeng, P. and He, Y. S. and Yang, H. M.},
  journal={Journal of Space Science and Experiment},
  year={2025},
  volume={2},
  number = {5},
  pages={116--124}
}

\end{document}